\documentclass{article}
\usepackage[utf8]{inputenc}
\usepackage[hidelinks]{hyperref}
\usepackage[a4paper,left=2.5cm, right=2.5cm,top=3.5cm, bottom=3.5cm]{geometry}
\usepackage{amsmath}
\usepackage{amssymb}
\usepackage{graphicx}
\usepackage[sort&compress,numbers]{natbib}

\usepackage{color}
\usepackage[makeroom]{cancel}
\usepackage[normalem]{ulem}
\newcommand{\KaTie}{\textsf{Ka\hspace{-0.4ex}T\hspace{-0.4ex}{\i}e}}
\newcommand{\LxJet}{\textsf{LxJet}}
\definecolor{pkcolor}{rgb}{0,0.1,0.7}

\newcommand\pkout{\marginpar{\color{pkcolor}$\clubsuit$}\bgroup\markoverwith{\color{pkcolor}{\rule[01ex]{2pt}{0.8pt}}}\ULon}

\definecolor{kkcolor}{rgb}{1,0,0}

\newcommand\kkout{\marginpar{\color{kkcolor}$\clubsuit$}\bgroup\markoverwith{\color{kkcolor}{\rule[04ex]{2pt}{0.8pt}}}\ULon}

\definecolor{mbcolor}{rgb}{0.3,0,1}

\newcommand\mbout{\marginpar{\color{mbcolor}$\clubsuit$}\bgroup\markoverwith{\color{mbcolor}{\rule[04ex]{2pt}{0.8pt}}}\ULon}

\definecolor{avhcolor}{rgb}{0,0.7,0.7}

\newcommand\avhout{\marginpar{\color{avhcolor}$\clubsuit$}\bgroup\markoverwith{\color{avhcolor}{\rule[04ex]{2pt}{0.8pt}}}\ULon}

\title{Forward trijet production in p-p and p-Pb collisions at LHC}
\author{Marcin Bury$^{a,b}$, Andreas van Hameren$^b$, Piotr Kotko$^c$, Krzysztof Kutak$^b$ 
\\ \\
$^a${\it  Institut f\"ur Theoretische Physik, Universit\"at Regensburg, }\\ 
    {\it D-93040 Regensburg, Germany} \\ \\
$^b$ {\it Institute of Nuclear Physics, Polish Academy of Sciences} \\
     {\it  Radzikowskiego 152, 31-342 Krak\'ow, Poland } \\ \\
$^c${\it AGH University Of Science and Technology, Physics Faculty,} \\ 
    {\it Mickiewicza 30, 30-059 Krak\'ow, Poland} 
}
\date{}

\begin{document}
\maketitle
\begin{abstract}

    We calculate various azimuthal angle distributions for three jets produced  in the forward rapidity region with transverse momenta $p_T>20\,\mathrm{GeV}$ in proton-proton (p-p) and proton-lead (p-Pb) collisions at center of mass energy $5.02\,\,\mathrm{TeV}$. We use the multi-parton extension of the so-called small-$x$ Improved Transverse Momentum Dependent factorization (ITMD). We study effects related to change from the standard $k_T$-factorization to ITMD factorization as well as changes as one goes from p-p collision to p-Pb. We observe rather large differences in the distribution when we change the factorization approach, which allows to both improve the small-$x$ TMD gluon distributions as well as validate and improve the factorization approach. We also see significant depletion of the nuclear modification ratio, indicating a possibility of searches for saturation effects using trijet final states in a more exclusive way than for dijets.
    
\end{abstract}

\section{Introduction}
So far, the complete trijet production in high energy hadron-hadron collisons with initial off-shell partons was  discussed only within the $k_T$-factorization approach \cite{vanHameren:2013fla,VanHaevermaet:2020rro}. 
Here we wish to discuss for the first time the production of three jets within the so-called \emph{small-$x$ Improved Transverse Momentum Dependent}  (ITMD) approach \cite{Kotko:2015ura} that accounts for gluon saturation effects, off-shell hard matrix elements, and involves several transverse momentum dependent (TMD) gluon distributions.
The trijet production process at LHC kinematics is of great interest, since, as follows from a recent study by three of us \cite{VanHaevermaet:2020rro}, it has great potential to uncover details of dynamics related to transverse momentum dependence of proton constituents and to test properties of parton showers. Furthermore it allows to study the features of ITMD factorization, and, last but not least, it constitutes the real emission contribution to dijet production at NLO.

Before we continue with three jets, let us first summarize the theoretical formalisms used to calculate forward jet production in proton-proton and proton-nucleus collisions. This will also allow to set up the terminology to avoid possible confusion.

There are two quite distinct factorization approaches that make use of the parton distribution functions depending on the transverse momentum. First are the leading power factorization theorems of QCD, usually called the \emph{TMD factorization theorems} \cite{Collins:2011zzd}. Because they hold to leading power in $k_T/\mu$ at fixed energy ($k_T$ being the transverse momentum of incoming partons and $\mu$ the hard scale), the partonic processes entering the factorization formulae are  calculated fully on-mass-shell, {\it i.e.}\ the transverse momentum $k_T$ of incoming partons is neglected in the hard part\footnote{There is also a Monte Carlo approach to evaluate TMD distributions called the Parton Branching method \cite{Hautmann:2017fcj,Martinez:2018jxt}}. It is not neglected, however, in the hadronic soft part. Thus the only dependence on incoming parton momenta is in the PDFs. The evolution equations for the TMD PDFs are the Collins-Soper-Sterman equations and they resum the large logarithms of $k_T/\mu$ \cite{Collins:1984kg}. 
On the other hand the \emph{$k_T$-factorization} also called the \emph{high energy factorization} (HEF) focuses on the small Bjorken $x$ limit, not neglecting the powers $k_T/\mu$ \cite{Catani:1990eg,Levin:1991ry,Collins:1991ty,Deak:2009xt}. Thus, the incoming partons (usually gluons, as they dominate at high energy) carry the transverse momentum and the partonic processes must be calculated off-shell. In this context, the TMD PDFs are also called unintegrated PDFs. The corresponding evolution equations typically resum the logarithms of energy (or equivalently  $1/x$) by means of the Balitsky-Fadin-Kuraev-Lipatov (BFKL) equation \cite{Kuraev:1977fs, Balitsky:1978ic} or its extensions, like for instance the  Catani-Ciafaloni-Fiorani-Marchesini (CCFM) \cite{Ciafaloni:1987ur,Catani:1989sg,Catani:1989yc} or approaches/models combining BFKL and DGLAP like  Kwieciński-Martin-Staśto (KMS) equation~\cite{Kwiecinski:1997ee} or Kimber-Martin-Ryskin-Watt \cite{Watt:2003vf} (KMRW).
For an approach based on application on BFKL amplitudes merged with the DGLAP parton densities see also \cite{Caporale:2016soq}.

The BFKL-type approach does not account for very large gluon densities inside hadrons -- the evolution equations are linear and the gluon densities can grow power-like with energy, eventually violating the unitarity bound. The QCD theory predicts however a nonlinear generalization of the BFKL equation --  the Balitsky-Kovchegov (BK) equation \cite{Balitsky:1995ub,Kovchegov:1999yj}, which exhibits \emph{gluon saturation}, $i.e.$ a state where almost all gluons have momenta $k_T\sim Q_s$, where $Q_s$ is the saturation scale. The BK equation is the mean field approximation to the more general system of equations known as 
the B-JIMWLK equations \cite{Balitsky:1995ub,
JalilianMarian:1997jx,JalilianMarian:1997gr,JalilianMarian:1997dw,Kovner:2000pt,Kovner:1999bj,Weigert:2000gi,Iancu:2000hn,Ferreiro:2001qy}, which describe evolution of various gluon operators supplemented with Wilson lines. These operators have very different behavior for small $k_T$ but coincide (or vanish very fast) at large $k_T\gg Q_s$, $i.e.$ in the linear regime. 

The modern effective theory incorporating saturation is the \emph{Color Glass Condensate} (CGC) theory (see e.g. \cite{Gelis:2010nm}; for a comprehensive pedagogical review of high energy QCD see \cite{Kovchegov:2012mbw}). In this theory, the gluon operators coupling to various particle production processes are averaged over random color sources of a dense target. It is important to underline two basic aspects of this approach: i) the eikonal approximation is assumed (recently also the non-eikonal corrections have been studied \cite{Agostini:2019avp}), ii) the gluon operators contain both the leading power contribution (leading twist) and subleading power corrections, containing in principle the genuine multi-parton operators. The latter lead to resummation of the multiple parton interactions (MPI). 

In jet phenomenology at LHC we typically assume the hard scale $\mu$ is set up by the average transverse momentum $p_T$ of jets. We consider here not so hard jets, with $p_T$ above, say, 15-20 GeV so that we are still sensitive to saturation, at sufficiently large energy and forward rapidity. In that regime, many simplifications occur. First, if one is interested in the back-to-back jet region only ($k_T\ll\mu$ assuming the collinear projectile), the leading power extraction \cite{Dominguez:2011wm,Altinoluk:2018byz} leads to an effective TMD factorization with on-shell partonic amplitudes and several small-$x$ leading power TMD gluon distributions containing various Wilson line operators ensuring gauge invariance and resumming collinear gluons \cite{Bomhof:2006dp}. In case we are interested in full description of jet imbalance (thus any $k_T$ between $Q_s$ and $\mu$), we can extend the above effective TMD formalism to incorporate the kinematic twist corrections. This is done via keeping the incoming gluon off-shell in the amplitude and assuring the gauge invariance \cite{vanHameren:2012uj,vanHameren:2012if,Kotko:2014aba}, which is equivalent to Lipatov's vertices in quasi-multi-Regge kinematics \cite{Antonov:2004hh}. Such formalism has been first developed for dijets in \cite{Kotko:2015ura} and is referred to as small-$x$ improved TMD factorization (ITMD). It is equivalent to CGC expressions for dilute-dense collisions  \cite{Dumitru:2005gt} with all kinematic twist corrections isolated and resummed, while neglecting the genuine twist corrections \cite{Altinoluk:2019fui}.
The kinematic twist corrections are the power corrections to the hard process, whereas the latter contributions come from hadronic matrix elements of more than two gluon field strength operators connected to the hard process. They are also called the genuine MPI contributions (note these MPI contributions are rather different than soft MPIs used in general purpose Monte Carlo generators). For more details consult  \cite{Altinoluk:2019fui,Altinoluk:2019wyu}.

The extension of the
ITMD formalism for three jets used in the present work can be constructed analogously, provided the TMD gluon distribution definitions are known. Among others, for that purpose the operator structures for three-and four-jet processes have been explicitly calculated in \cite{Bury:2018kvg}. 
The precise factorization formula shall be given in the next section. In the end, let us note that at present, the ITMD formalism does not account for the  linearly-polarized gluons in unpolarized target. 
In CGC theory, such a contribution is absent for massless two-particle
production, but appears in heavy quark production  \cite{Marquet:2017xwy} and will appear in higher multiplicity processes as it has been already observed in the correlation limit 
for three-parton final state \cite{Altinoluk:2020qet} basing on the quark-initiated three-jet production formulae in CGC \cite{Iancu:2018hwa}.

Therefore
in what follows we shall call our extension of the ITMD formalism to
multipartonic processes ITMD*, to indicate that it does not take linearly polarized gluons into account yet. The construction of the full ITMD framework is
left for the future.
\section{ITMD* for three jets}

The generic formula for multiparticle production within the ITMD* approach has been given in \cite{Bury:2018kvg} in terms of color-ordered amplitudes (see e.g. \cite{Mangano:1990by} for a review of the color decomposition technique). For a specific case of forward particle production, where a dilute proton $p$ (probed at large $x$) collides with a dense target $A$ (probed at small $x$) the generic formula reads:
\begin{multline}
d\sigma_{pA\rightarrow n}=
\int\! \frac{dx_1}{x_1} \frac{dx_2}{x_2}\int\! d^2k_T 
\int\! \frac{d\Gamma_n}{2\hat s}\, \sum_{a} \sum_{b_1,\dots,b_n}  
x_1f_{a/p}(x_1,\mu)\,\, \\
\times \vec{\mathcal{A}}^{\,\dagger}_{ag\rightarrow b_1\dots b_n} 
\left\{\boldsymbol{C}\circ \mathbf{\Phi}_{ag\rightarrow b_1\dots b_n}\left(x_2,k_T\right)\right\} 
\vec{\mathcal{A}}_{ag\rightarrow b_1\dots b_n}\,\,,
\label{eq:xsect}
\end{multline}
where $d\Gamma_n$ is the $n$-particle phase space, $f_{a/p}$ the collinear PDF for parton $a$, depending on longitudinal momentum fraction $x_1$ and factorization scale $\mu$,  $b_1,\dots,b_n$ are various final state partons contributing to $ag\rightarrow b_1\dots b_n$ partonic sub-process. 
Further  $\vec{\mathcal{A}}$ is a vector of tree-level color-ordered amplitudes for given partonic sub-process, $\boldsymbol{C}$ is the color matrix and the symbol $\circ$ is the Hadamard (element-wise)
multiplication, $\left(\boldsymbol{A}\circ\boldsymbol{B}\right)_{ij}=\boldsymbol{A}_{ij}\boldsymbol{B}_{ij}$. Finally the $ \mathbf{\Phi}$ is the matrix of unpolarized TMD gluon distributions in the color-ordered basis. Entries of this matrix consist in linear combinations of the basis TMD gluon distributions given below. For three and four jets they have been explicitly calculated in \cite{Bury:2018kvg}. It has to be noted that the formalism is not restricted to a specific color representation, and explicit formulas for a particular one are given in Appendix~\ref{AppendixA}.
The operator definitions of the basic unpolarized TMD gluon distributions are\footnote{We assume that the correlators are real and we give only one of the two possible forms.}:
\begin{equation}
\mathcal{F}_{qg}^{(1)}\left(x,k_{T}\right)  =\mathrm{F.T.}\,\left<\mathrm{Tr}\left[\hat{F}^{i+}\left(\xi\right)\mathcal{U}^{[-]\dagger}\hat{F}^{i+}\left(0\right)\mathcal{U}^{[+]}\right]\right>\, ,
\label{eq:Fqg1}
\end{equation}
\begin{equation}
\mathcal{F}_{qg}^{(2)}\left(x,k_{T}\right)  =\mathrm{F.T.}\,\left<\frac{\mathrm{Tr}\left[\mathcal{U}^{[\square]}\right]}{N_{c}}\mathrm{Tr}\left[\hat{F}^{i+}\left(\xi\right)\mathcal{U}^{[+]\dagger}\hat{F}^{i+}\left(0\right)\mathcal{U}^{[+]}\right]\right>\, ,
\end{equation}
\begin{equation}
\mathcal{F}_{qg}^{(3)}\left(x,k_{T}\right)  =\mathrm{F.T.}\,\left<\mathrm{Tr}\left[\hat{F}^{i+}\left(\xi\right)\mathcal{U}^{[+]\dagger}\hat{F}^{i+}\left(0\right)\mathcal{U}^{[\square]}\mathcal{U}^{[+]}\right]\right>\, ,
\label{eq:Fqg3}
\end{equation}
\begin{equation}
\mathcal{F}_{gg}^{(1)}\left(x,k_{T}\right)  =\mathrm{F.T.}\,\left<\frac{\mathrm{Tr}\left[\mathcal{U}^{[\square]\dagger}\right]}{N_{c}}\mathrm{Tr}\left[\hat{F}^{i+}\left(\xi\right)\mathcal{U}^{[-]\dagger}\hat{F}^{i+}\left(0\right)\mathcal{U}^{[+]}\right]\right>\, ,
\label{eq:Fgg1}
\end{equation}
\begin{equation}
\mathcal{F}_{gg}^{(2)}\left(x,k_{T}\right)  =\mathrm{F.T.}\,\frac{1}{N_{c}}\left<\mathrm{Tr}\left[\hat{F}^{i+}\left(\xi\right)\mathcal{U}^{[\square]\dagger}\right]\mathrm{Tr}\left[\hat{F}^{i+}\left(0\right)\mathcal{U}^{[\square]}\right]\right>\, ,
\label{eq:Fgg2}
\end{equation}
\begin{equation}
\mathcal{F}_{gg}^{(3)}\left(x,k_{T}\right)=\mathrm{F.T.}\,\left<\mathrm{Tr}\left[\hat{F}^{i+}\left(\xi\right)\mathcal{U}^{[+]\dagger}\hat{F}^{i+}\left(0\right)\mathcal{U}^{[+]}\right]\right>\,,
\end{equation}
\begin{equation}
\mathcal{F}_{gg}^{(4)}\left(x,k_{T}\right)=\mathrm{F.T.}\,\left<\mathrm{Tr}\left[\hat{F}^{i+}\left(\xi\right)\mathcal{U}^{[-]\dagger}\hat{F}^{i+}\left(0\right)\mathcal{U}^{[-]}\right]\right>\,,
\end{equation}
\begin{equation}
\mathcal{F}_{gg}^{(5)}\left(x,k_{T}\right)=\mathrm{F.T.}\,\left<\mathrm{Tr}\left[\hat{F}^{i+}\left(\xi\right)\mathcal{U}^{[\square]\dagger}\mathcal{U}^{[+]\dagger}\hat{F}^{i+}\left(0\right)\mathcal{U}^{[\square]}\mathcal{U}^{[+]}\right]\right>\,,
\end{equation}
\begin{equation}
\mathcal{F}_{gg}^{(6)}\left(x,k_{T}\right)=\mathrm{F.T.}\,\left<\frac{\mathrm{Tr}\left[\mathcal{U}^{[\square]}\right]}{N_{c}}\frac{\mathrm{Tr}\left[\mathcal{U}^{[\square]\dagger}\right]}{N_{c}}\mathrm{Tr}\left[\hat{F}^{i+}\left(\xi\right)\mathcal{U}^{[+]\dagger}\hat{F}^{i+}\left(0\right)\mathcal{U}^{[+]}\right]\right>\,,\label{eq:Fgg6}
\end{equation}
\begin{equation}
\mathcal{F}_{gg}^{(7)}\left(x,k_{T}\right)=\mathrm{F.T.}\,\left<\frac{\mathrm{Tr}\left[\mathcal{U}^{[\square]}\right]}{N_{c}}\mathrm{Tr}\left[\hat{F}^{i+}\left(\xi\right)\mathcal{U}^{[\square]\dagger}\mathcal{U}^{[+]\dagger}\hat{F}^{i+}\left(0\right)\mathcal{U}^{[+]}\right]\right>\, ,
\label{eq:Fgg7}
\end{equation}
where $\mathrm{F.T.}$ stands for the Fourier transform
\begin{equation}
\mathrm{F.T.} = 2\int\frac{d\xi^{-}d^{2}\xi_{T}}{\left(2\pi\right)^{3}P^{+}}\,e^{ixP^{+}\xi^{-}-i\vec{k}_{T}\cdot\vec{\xi}_{T}}\, .
\end{equation}
The angle brackets represent the hadronic matrix element $\left<\dots\right>=\left<P\right|\dots\left|P\right>$, where $P$ is the momentum of the hadron. 
Further $\hat{F}^{\mu\nu}=F_{a}^{\mu\nu}T^{a}$, where $T^{a}$ are the color generators. We employ standard light-cone basis, with hadron traveling along the 'plus' direction. The fields are separated in the light-cone 'minus' and transverse directions:
\begin{equation}
    \xi=\left(\xi^{+}=0,\xi^{-},\vec{\xi}_{T}\right)
\end{equation}
The two staple-like fundamental representation Wilson lines connecting the fields are
\begin{multline}
\mathcal{U}^{\left[\pm\right]}=\left[\left(0^{+},0^{-},\vec{0}_{T}\right),\left(0^{+},\pm\infty^{-},\vec{0}_{T}\right)\right]
\left[\left(0^{+},\pm\infty^{-},\vec{0}_{T}\right),\left(0^{+},\pm\infty^{-},\vec{\xi}_{T}\right)\right]\\
\times \left[\left(0^{+},\pm\infty^{-},\vec{\xi}_{T}\right),\left(0^{+},\xi^{-},\vec{\xi}_{T}\right)\right]\,,
\label{eq:staple}
\end{multline}
where the square brackets are the straight segments of the Wilson link. The Wilson loop is just two staples glued together:
\begin{equation}
\mathcal{U}^{\left[\square\right]}=\mathcal{U}^{\left[-\right]\dagger}\mathcal{U}^{\left[+\right]}\,.\label{eq:WilsonLoopDef}
\end{equation}

TMD matrices for the trijet case for two different color representations can be found in  \cite{Bury:2018kvg} and in Appendix~\ref{AppendixA}. The  gauge invariant color ordered amplitudes with off-shell inital state gluon can be calculated automatically at tree-level using the methods of \cite{vanHameren:2012uj,vanHameren:2012if,Kotko:2014aba}. In the following work we use two latter methods independently to cross check the results. 
They are independently implemented in two different Monte Carlo programs generating weighted or unweighted events according to the formula \ref{eq:xsect}: \KaTie~\cite{vanHameren:2016kkz} and \LxJet~\cite{Kotko_LxJet}.

Let us note that the ITMD formalism uses the small-$x$ limit of the basis TMD gluon distributions.  In that limit they can be rewritten in terms of matrix elements of CGC-style infinite Wilson lines with fixed transverse positions. Indeed, in the limit $x\rightarrow 0$ only transverse position survives in the definitions (\ref{eq:Fqg1})-(\ref{eq:Fgg7})  and the $x$ dependence emerges from the evolution in energy. Whilst the full evolution equation adequate for moderate and small $x$ for all correlators is not known (see \cite{Balitsky:2015qba,Balitsky:2016dgz} for initial attempts), the high energy limit is well controlled by the B-JIMWLK equations. Since B-JIMWLK is an evolution of the functional representing random color configurations in target it can be used for any operator. Assuming common initial distribution for operators contributing to dijet production, the proof of principle was given in \cite{Marquet:2016cgx,Marquet:2017xwy}. This type of calculation can be carried keeping the subleading $1/N_c$ corrections, but so far no distributions have been produced that incorporate data driven input. 

In the following work we follow another path, first employed in \cite{vanHameren:2016ftb}. We are going to use the TMD distribution appearing in the inclusive DIS processes, the so-called dipole distribution (\ref{eq:Fqg1}). In particular, we shall use the TMD coming from the BK equation  supplemented with subleading corrections following the KMS framework \cite{Kutak:2003bd} and fitted to $F_2$ data \cite{Kutak:2012rf}. This equation is actually more suitable for harder jets because of including the DGLAP and kinematical constraint contributions. 
Having the dipole gluon distribution, all other distributions appearing in the dijet production: $\mathcal{F}_{qg}^{(2)},\mathcal{F}_{gg}^{(1)},\mathcal{F}_{gg}^{(2)},\mathcal{F}_{qg}^{(3)}$ can be calculated in the mean field approximation often used in CGC theory and to leading number of colors, see \cite{vanHameren:2016ftb} for details.

In the above setup, $i.e.$ to leading number of colors and in the mean field approximation for the distribution of color sources in the target, the cross section for trijet production can be calculated using the same basis TMD distributions as for the dijet case.

\section{Numerical results}
\label{sec:numerics}
Before we present our results for the cross section, let us first discuss in more detail the basic TMD gluon distributions  $\mathcal{F}_{qg}^{(2)}$, $\mathcal{F}_{gg}^{(1)}$, $\mathcal{F}_{gg}^{(2)}$, $\mathcal{F}_{gg}^{(6)}$ calculated in \cite{vanHameren:2016ftb} from the Kutak-Sapeta (KS) dipole gluon distribution $\mathcal{F}_{qg}^{(1)}$, as well as the Weizsäcker-Williams gluon distribution $\mathcal{F}_{gg}^{(3)}$ calculated in  \cite{Kotko:2017oxg}. 
This will be necessary to properly interpret the results for the cross section, as the trijet topology probes the kinematic range so far unexplored in inclusive and dijet calculations.

\begin{figure}
\begin{center}
\includegraphics[width=6.cm]{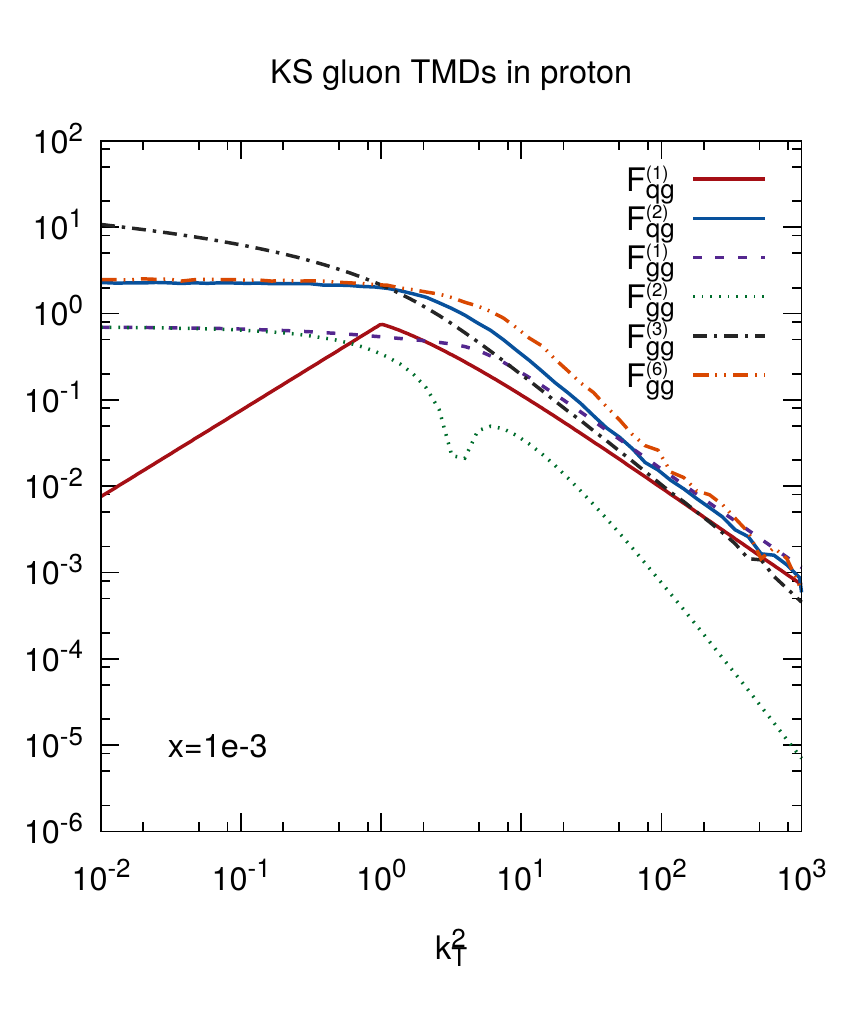}
\includegraphics[width=6.cm]{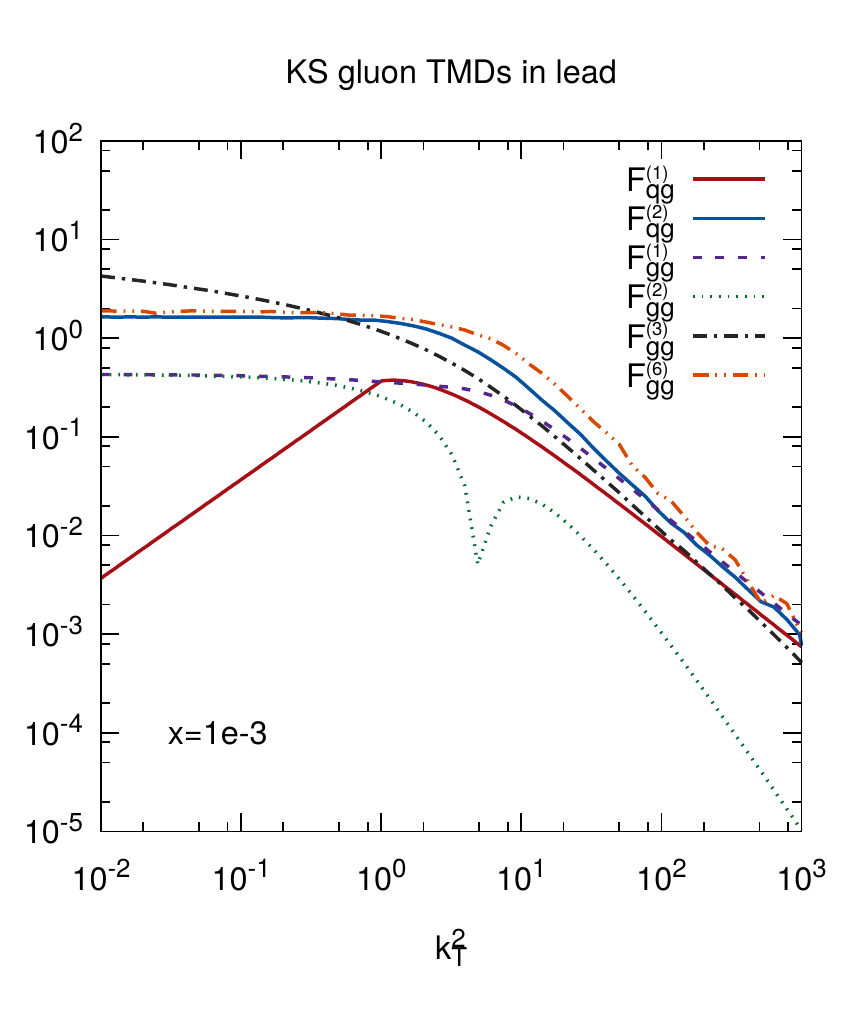}
\includegraphics[width=6.cm]{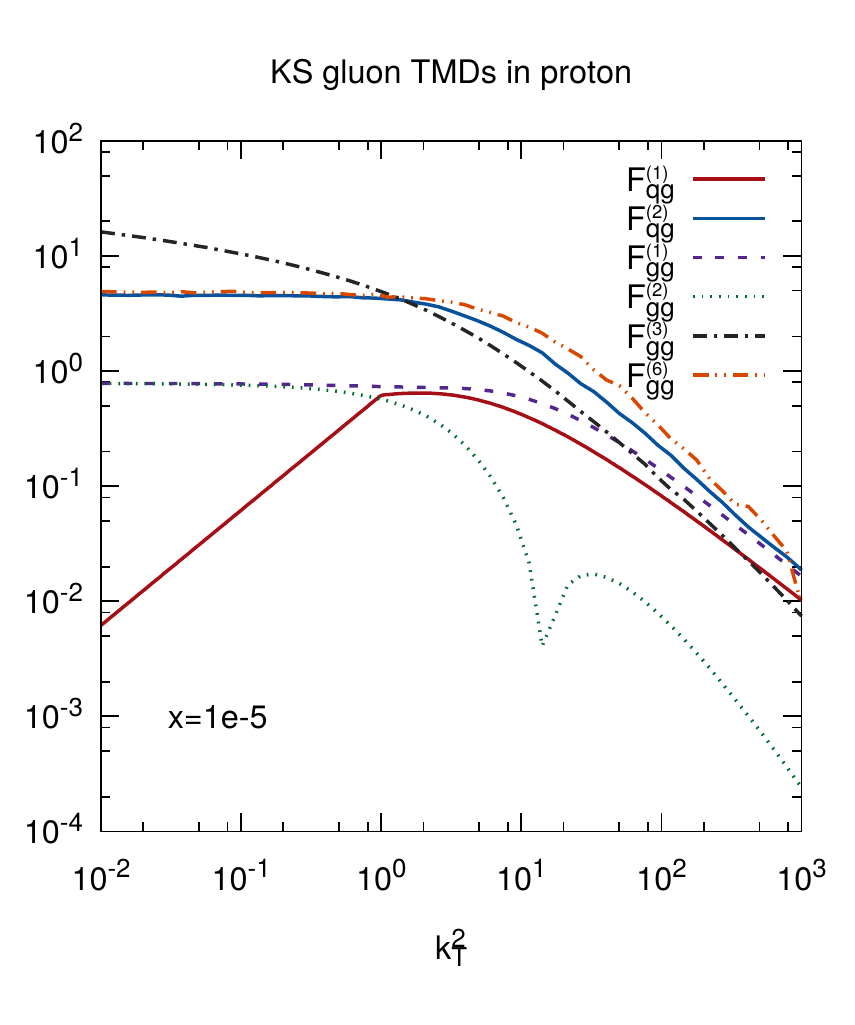}
\includegraphics[width=6.cm]{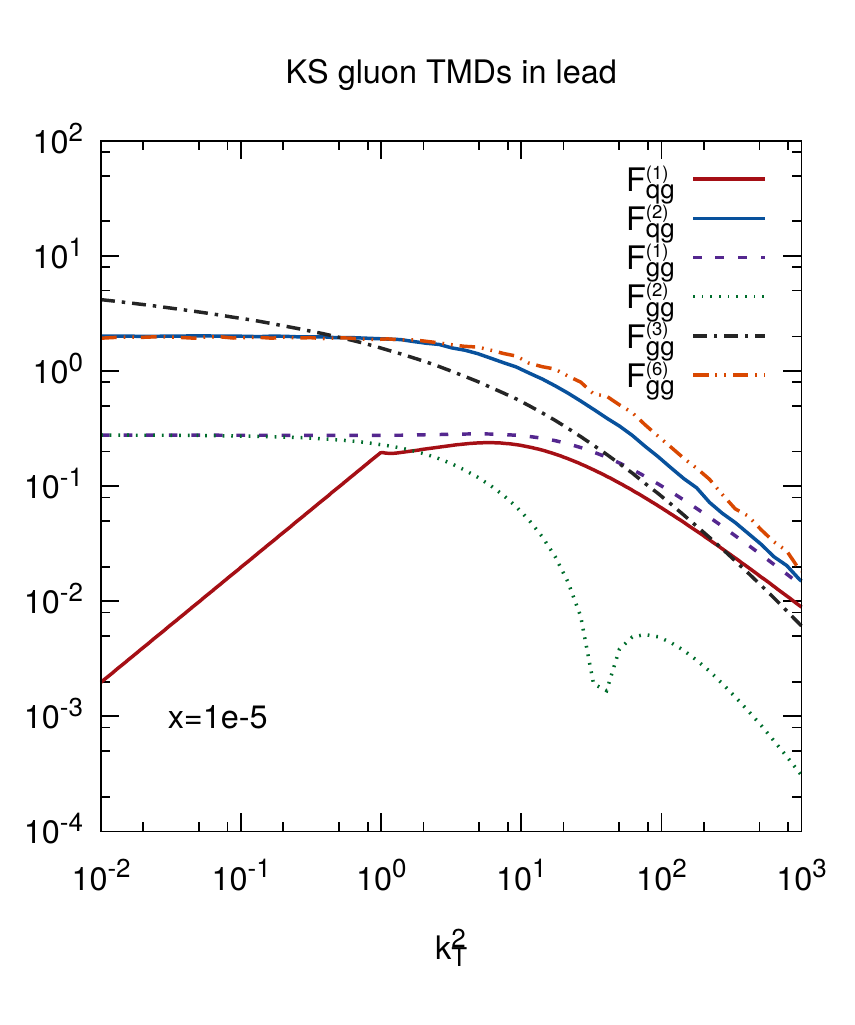}
\end{center}
 \caption{\label{fig:KSTMDx}The KS TMD gluon distributions for p and Pb at $x=0.001$ (top) and $x=0.00001$ (bottom).}
\end{figure}

\begin{figure}
\begin{center}
\includegraphics[width=6.1cm]{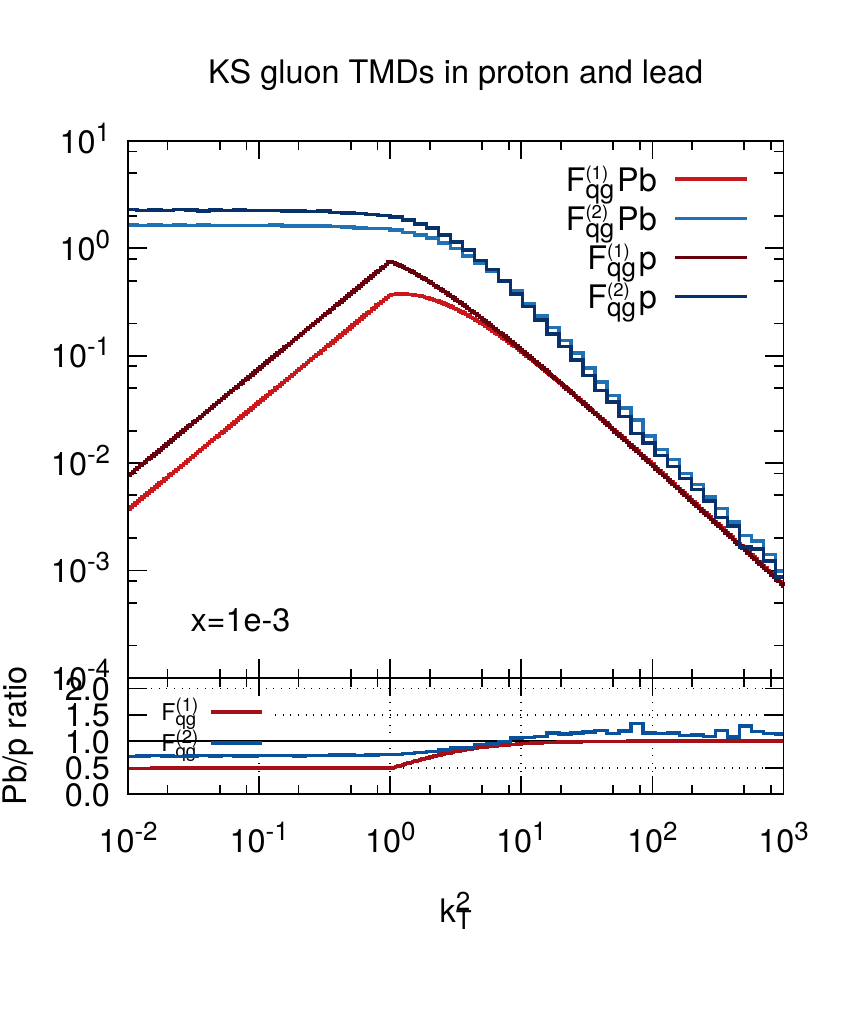}
\includegraphics[width=6.1cm]{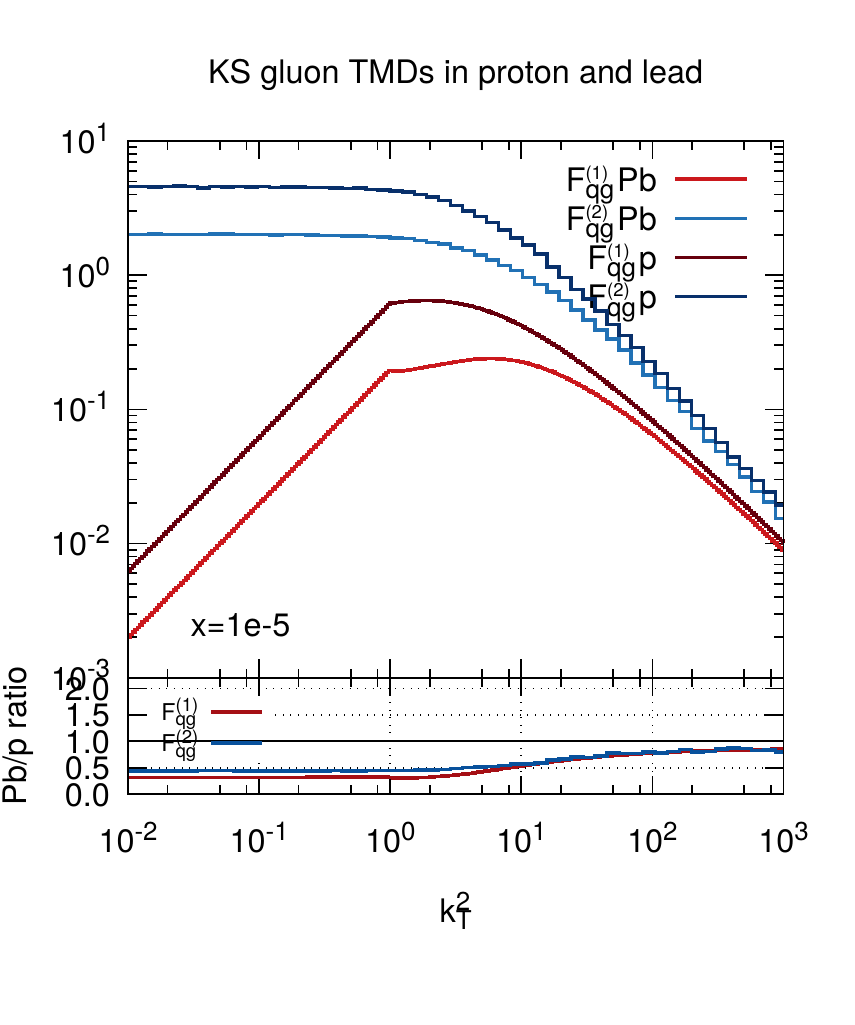}\\[-2ex]
\includegraphics[width=6.1cm]{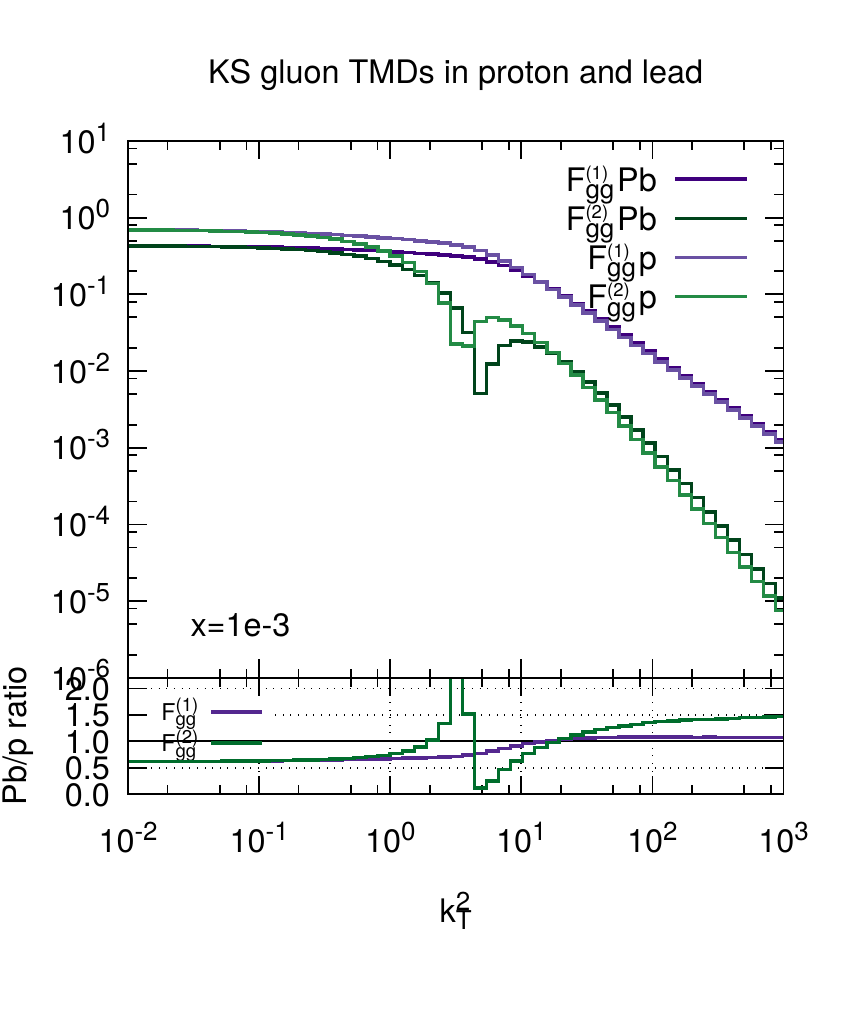}
\includegraphics[width=6.1cm]{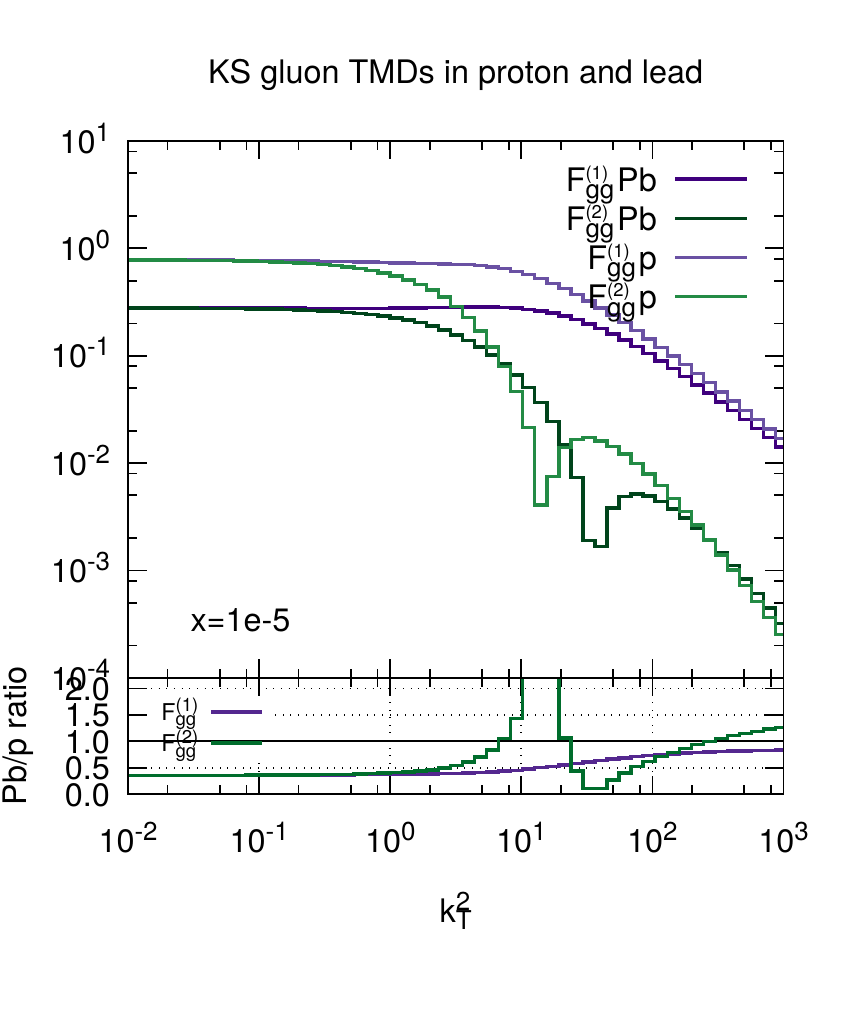}\\[-2ex]
\includegraphics[width=6.1cm]{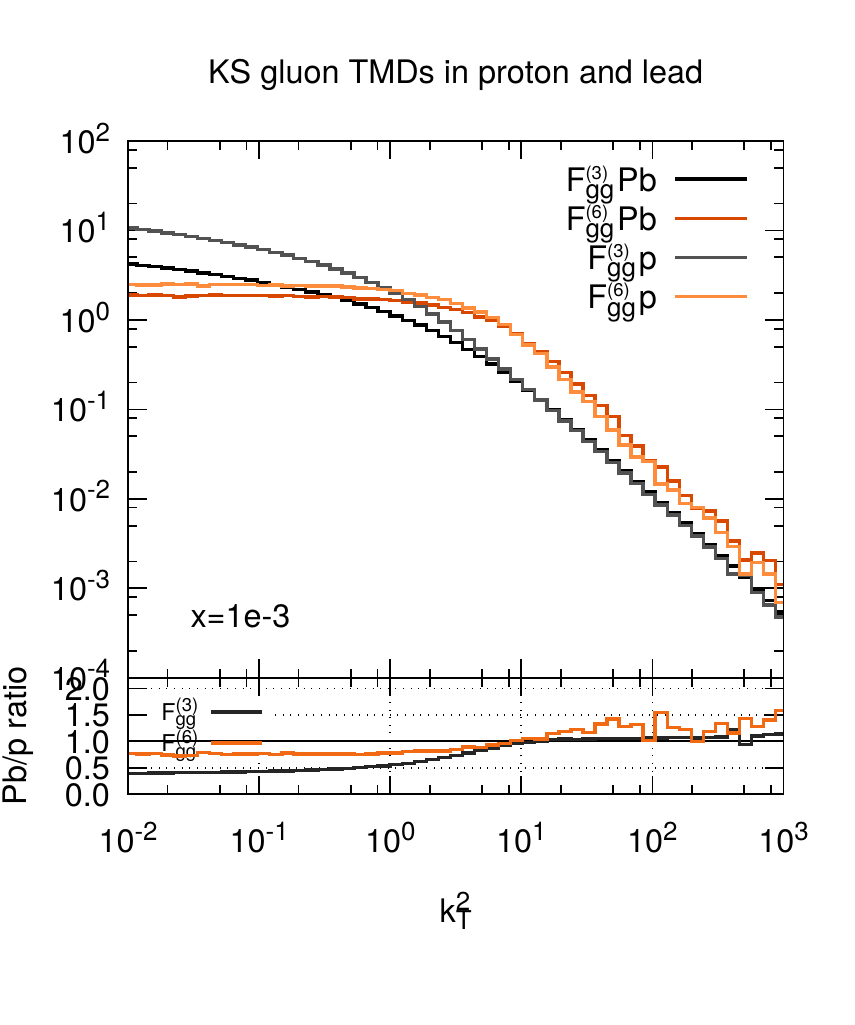}
\includegraphics[width=6.1cm]{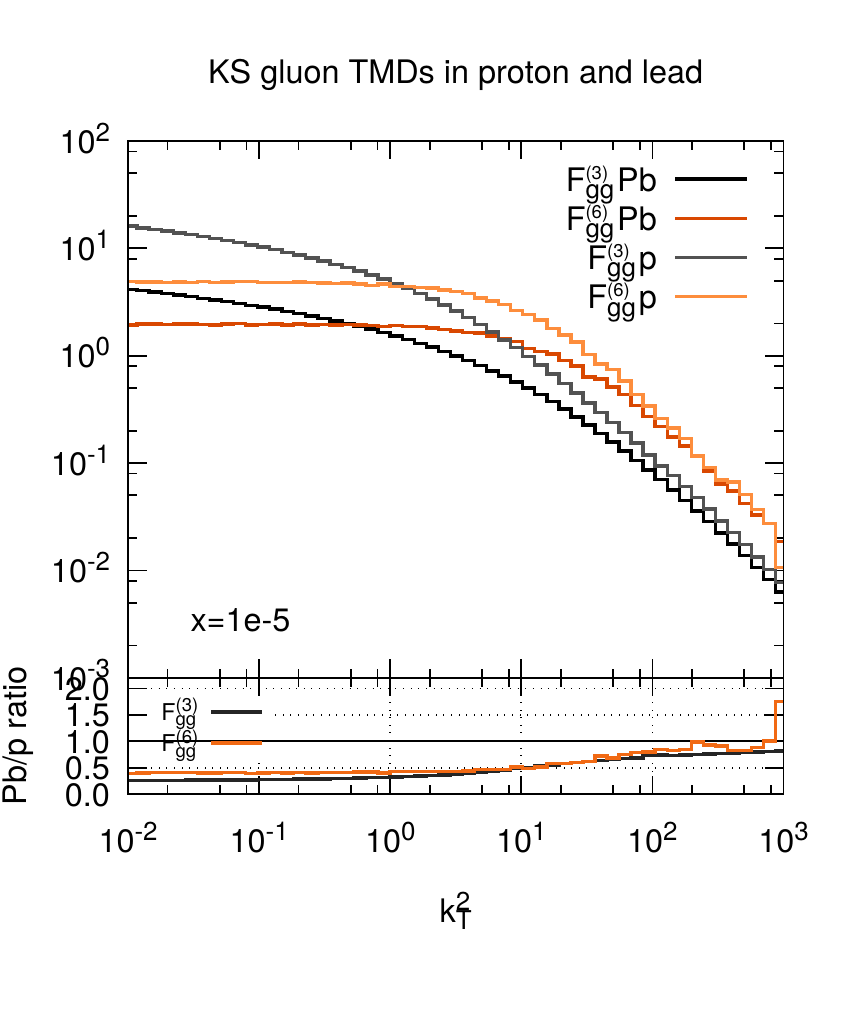}
\end{center}
\vspace{-4ex}
 \caption{\label{fig:Fqg_ratios}Comparison of KS TMDs for p and Pb and the Pb/p ratios for $x=0.001$ (left column) and $x=0.00001$ (right column).}
\end{figure}

The KS dipole gluon distribution $\mathcal{F}_{qg}^{(1)}$, has two trends: the dependence on $k_T$ below $\sim 1\,\mathrm{GeV}$ was approximated by a power-like falloff, whereas above that threshold the TMD is given by the solution of BK equation with subleading corrections.
The saturation of the distribution is visible as the clear smooth maximum developing for  $k_T \gtrsim 1\,\mathrm{GeV}$ and moving towards larger $k_T$ with decrease of $x$ (solid red line in Fig.~\ref{fig:KSTMDx}).
The remaining gluon distributions obtained from $\mathcal{F}_{qg}^{(1)}$  have universal behavior at large $k_T$ -- they decay like $\sim 1/k_T$. The exception is the $F_{gg}^{(2)}$ which decays much faster, so that it does not contribute to the perturbative tail $1/k_T$. 
Further, it becomes negative for some values of $k_T$ and approaches zero from below. Therefore, we plot its absolute value on the logarithmic scale in Fig.~\ref{fig:KSTMDx}. 
To see in greater detail the differences as we go from proton to lead we  plot also the ratios of gluon densities, Fig.~\ref{fig:Fqg_ratios}. Interestingly, whereas at very low $x$ all the distributions in lead are suppressed as compared to the proton, for moderate $x$ and large $k_T$ the ratios exceed one. 
The interpretation of this result is not obvious. 
One possible explanation is that for a given $x$, the gluon distribution in lead can be significantly broader compared to proton, and thus create an enhancement for large $k_T$, being still suppressed for small $k_T$ values.

Now we  are ready to discuss our present results for the trijet cross section. The processes that we are after are the  p-p and p-Pb collisions at $\sqrt{s}=5.02\,\mathrm{TeV}$ per nucleon and with demand that the three jets are produced in the forward rapidity window $3.2<|y^*_1,y^*_2,y^*_3|<4.9$, where $y^*$ in the rapidity defined in the CM frame. The jets are defined as the outgoing on-shell partons satisfying the $\Delta\phi-\Delta\eta$ cut with the radius $R=0.5$. We demand the jets have transverse momentum of at least $p_T>20\,\mathrm{GeV}$ for all jets. We order the jets according to their $p_T$, so that we can distinguish the leading, the sub-leading, and the soft jet: $p_{T1}>p_{T2}>p_{T3}$.
We put the factorization/renormalization scale equal to $(p_{T1}+p_{T2}+p_{T3})/3$, and shaded areas in plots represent a variation of this scale between a factor $1/2$ and $2$.

We perform calculation with the ITMD* framework described in the preceding section. For the collinear parton distributions we use the CTEQ10NLO set \cite{Lai:2010vv} obtained from LHAPDF6 \cite{Buckley:2014ana}, and the Kutak-Sapeta (KS) dipole gluon distribution \cite{Kutak:2012rf} to get the five TMD gluon distributions \cite{vanHameren:2016ftb} needed for the ITMD* at leading number of colors and in the mean field approximation, as discussed in details in the previous Section and above.
We consider the following partonic channels, for 5 flavors of quarks:
\begin{gather}
    g^*q \rightarrow qgg \, , \qquad 
    g^*q \rightarrow qq\bar{q} \, , \qquad 
    g^*q \rightarrow qq'\bar{q}' \\
    g^*g \rightarrow ggg \, , \qquad 
    g^*g \rightarrow gq\bar{q}\, ,
\end{gather}
where $q$ and $q'$ are quarks with necessarily different flavor. We do not include the sub-process with incoming anti-quark as it gives negligible contribution in the forward jet production case. The off-shell gauge invariant amplitudes were obtained numerically for fixed helicity and summed over on the event-by-event basis.

In the following we are interested in the azimuthal angle distributions of the jets. Thus, we consider the differential cross sections as a function of: (i) the azimuthal angle between the leading and sub-leading jets $\Delta\phi_{12}$, (ii) the azimuthal angle between the leading and the soft jet $\Delta\phi_{13}$, (iii) the azimuthal angle between the plane spanned by the two leading jets and the soft jet $\Delta\phi_{(12)3}$.

\begin{figure}
\begin{minipage}{14.5cm}
\begin{center}
\includegraphics[width=7cm]{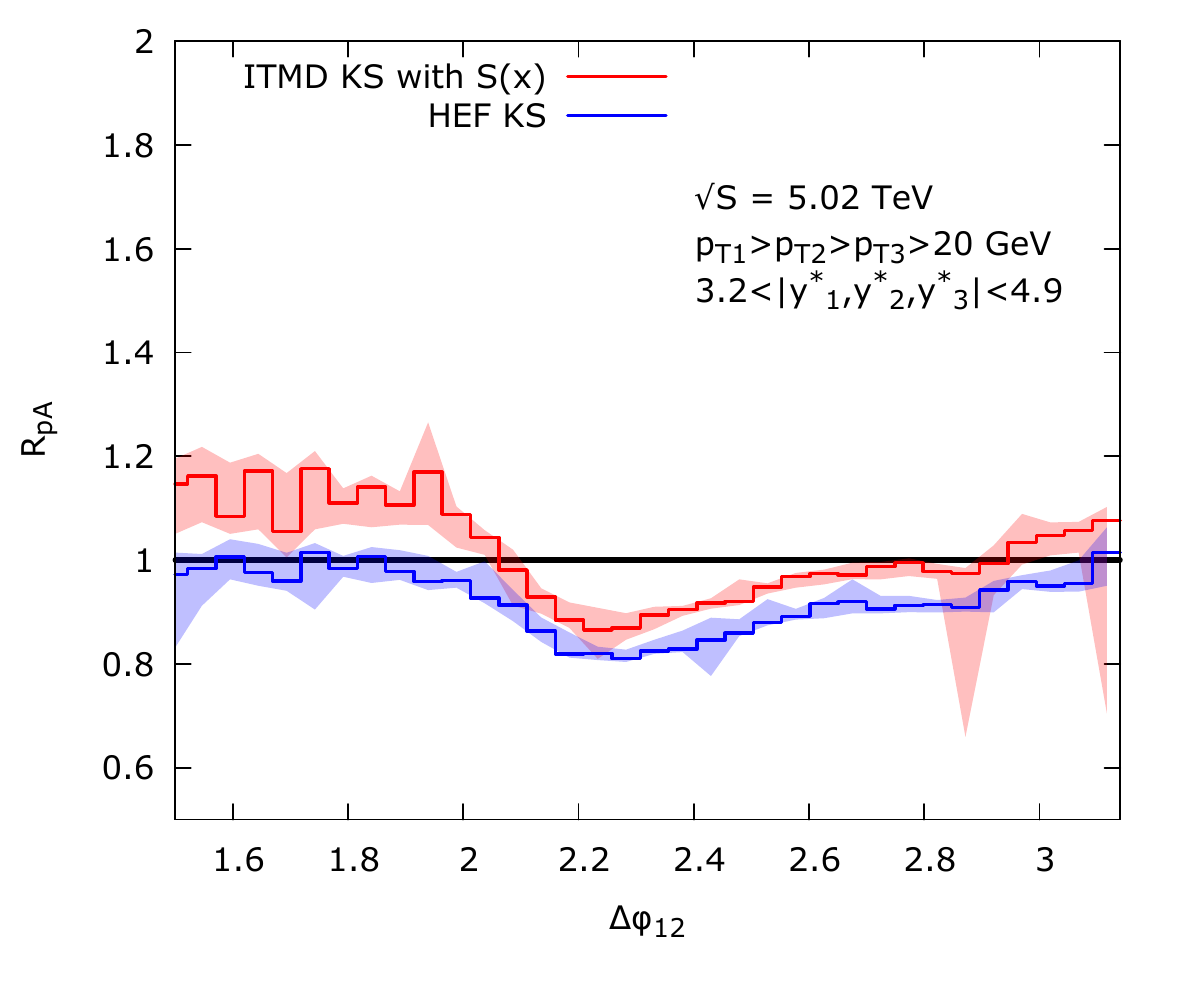}
\includegraphics[width=7cm]{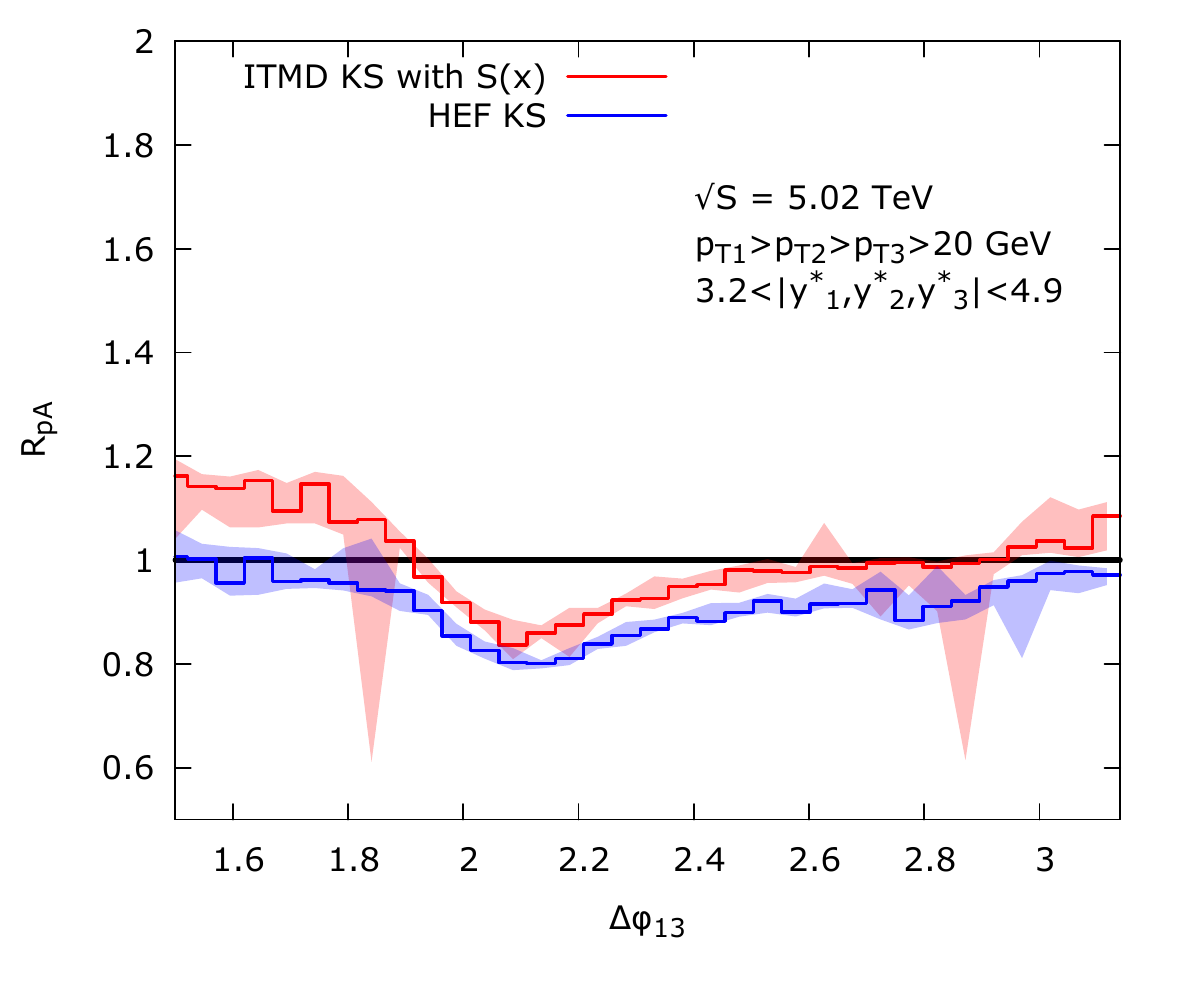}\\
\parbox[c]{7cm}{\includegraphics[width=7cm]{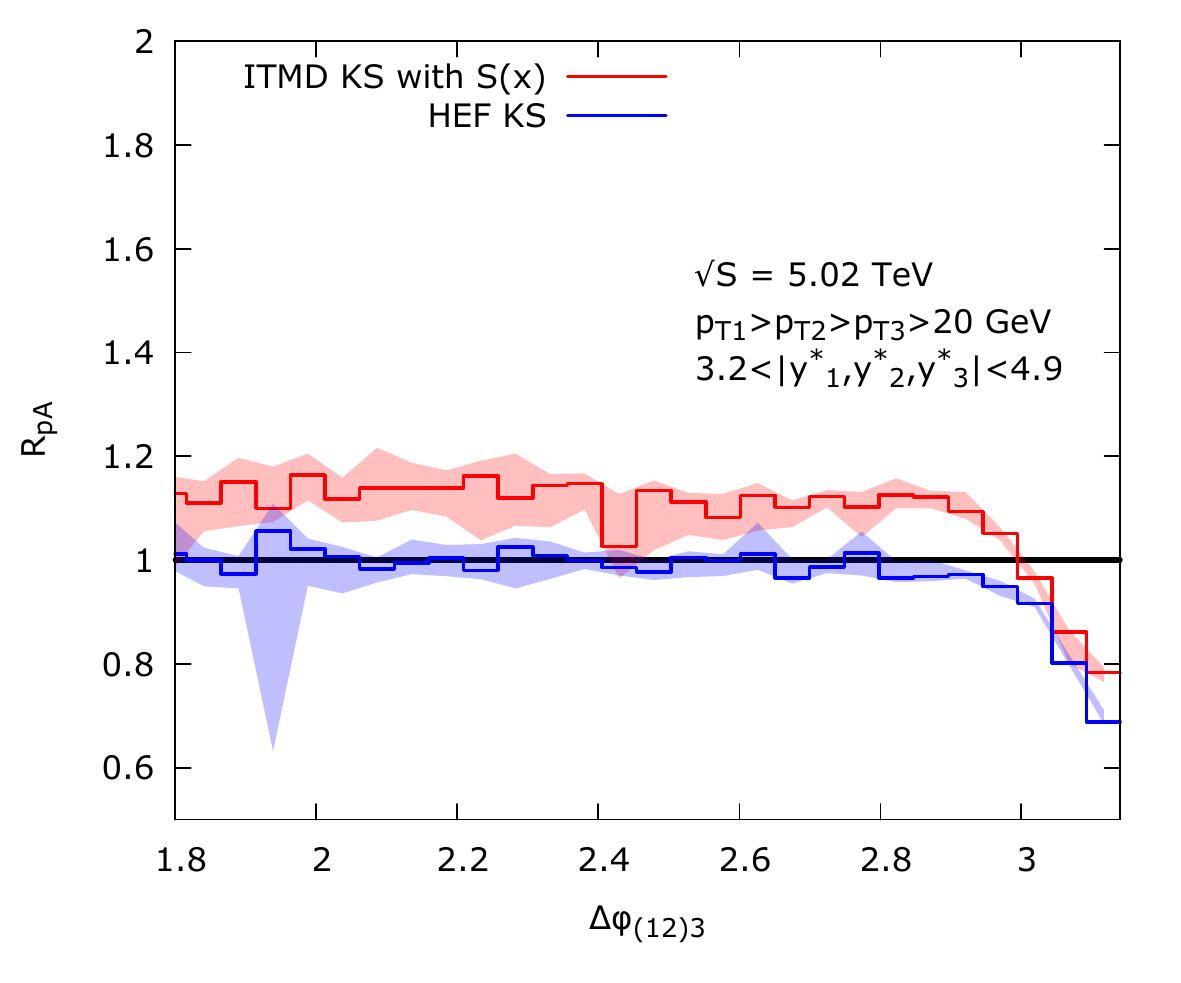}}
\hspace{0.5cm}\parbox[c]{6.3cm}{
   \caption{\label{fig:RpA}Nuclear modification ratio in azimuthal differences $\Delta\phi_{12}$, $\Delta\phi_{13}$ and  $\Delta\phi_{(12)3}$. The ITMD* calculation predicts less suppression in the corresponding back-to-back regions comparing to HEF and displays up to 10\% of an enhancement of the p-Pb cross section away from the correlation limit (for the possible origin of this effect see the discussion in the main text).}
   }
\end{center}  
\end{minipage}
\end{figure}

In addition to the absolute differential cross sections, a very useful observable  is the nuclear modification ratio that quantifies saturation effects. It is defined generically as
\begin{equation}
    R_{\mathrm{pPb}}=\frac{\frac{d\sigma^{p+Pb}}{d {\cal O}   }}{A\frac{d\sigma^{p+p}}{d {\cal O}   }}\,, 
\end{equation}
where the numerator corresponds to an observable in $pA$ collision and the denominator to an observable in $pp$ collisions, scaled by number of nucleons $A$. 
The deviation from unity suggests emergence of novel effects as one goes from $pp$ to $pA$. In our case it reflects the emergence of nonlinearities leading to the gluon saturation 
(for $R<1$) and possibly to anti-shadowing (for $R>1$) due to the broadening of the TMD gluon distributions used in the calculations.
In Fig.~\ref{fig:RpA} we plot nuclear modification ratio as a function of $\Delta\phi_{12}$, $\Delta\phi_{13}$ and  $\Delta\phi_{(12)3}$. 
We consider the following scenarios:
\begin{itemize}
    \item 
    The ITMD* case with the KS TMD gluon distributions with  the $x$-dependent nuclear target area $S(x)$. 
    This factor enters the calculation of the TMD distributions as follows.  
    The dipole KS gluon density is integrated over the impact parameter. However, the procedure to get the rest of the TMD gluon distributions requires dividing out by nuclear target area $S$.
    Inclusion of $x$ dependence, $S=S(x)$, guarantees that the normalisation of a dipole cross section reaches unity for large dipoles, as expected in the black disk limit. 
    Without the $x$ dependence the unitarity is not guaranteed since the higher order corrections are applied in the KS approach only to linear part of the BK equation. 
    \item The HEF case with nonlinear and linear KS dipole gluon distribution. 
    The latter is the exact dilute limit (i.e. purely linear limit) of the ITMD* factorization and the CGC formalism (denoted as KS-lin in the plots).
\end{itemize}
 Comparison of the above cases allows us to quantify the role of gauge links by comparing ITMD* with HEF when the gauge 
 links are neglected but the nonlinearity is kept, as well as to quantify the combined effect of gauge links  
 and nonlinearities by comparing ITMD* to completely dilute limit.

\begin{figure}
\begin{minipage}{14.5cm}
\begin{center}
\includegraphics[width=7cm]{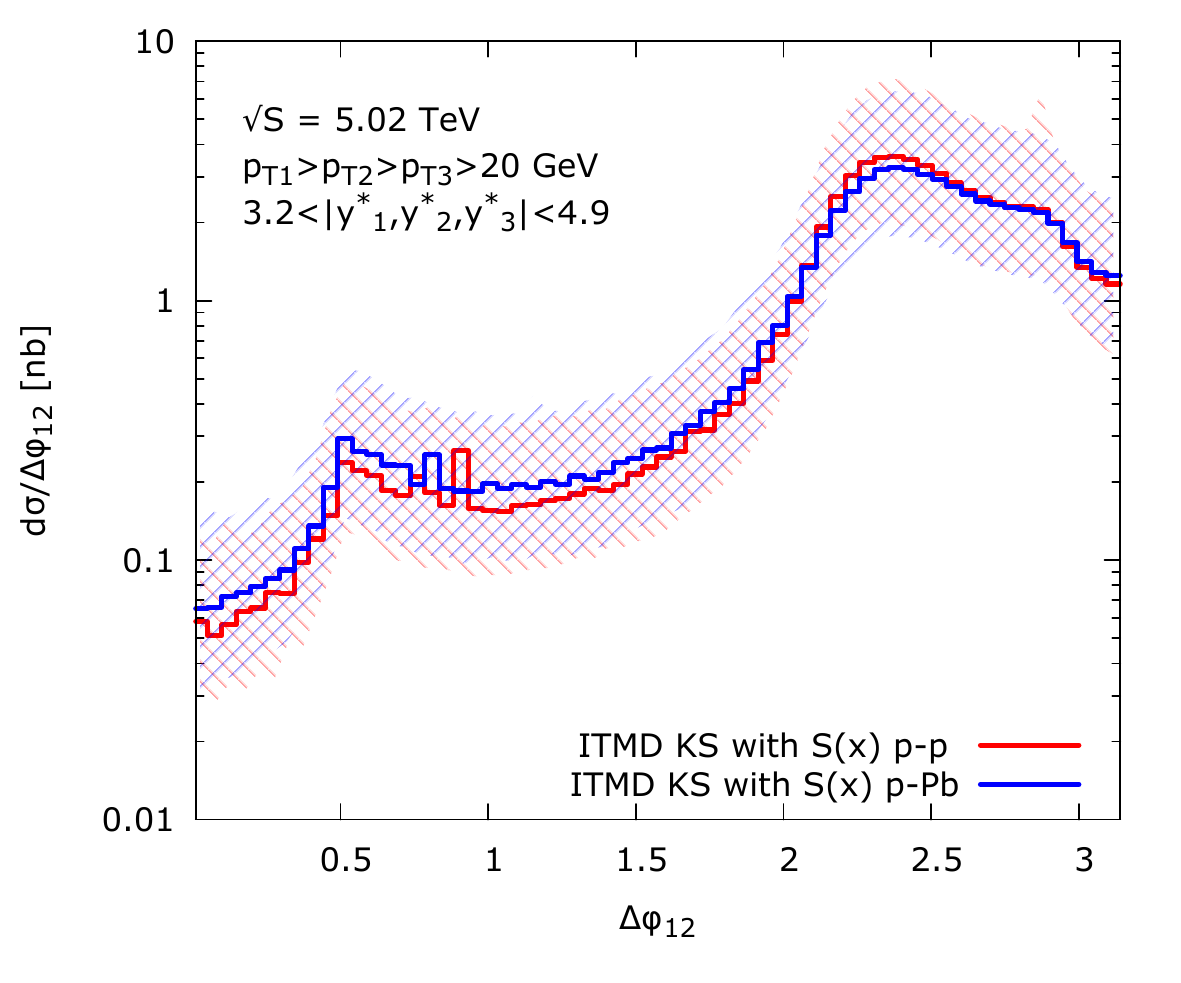}
\includegraphics[width=7cm]{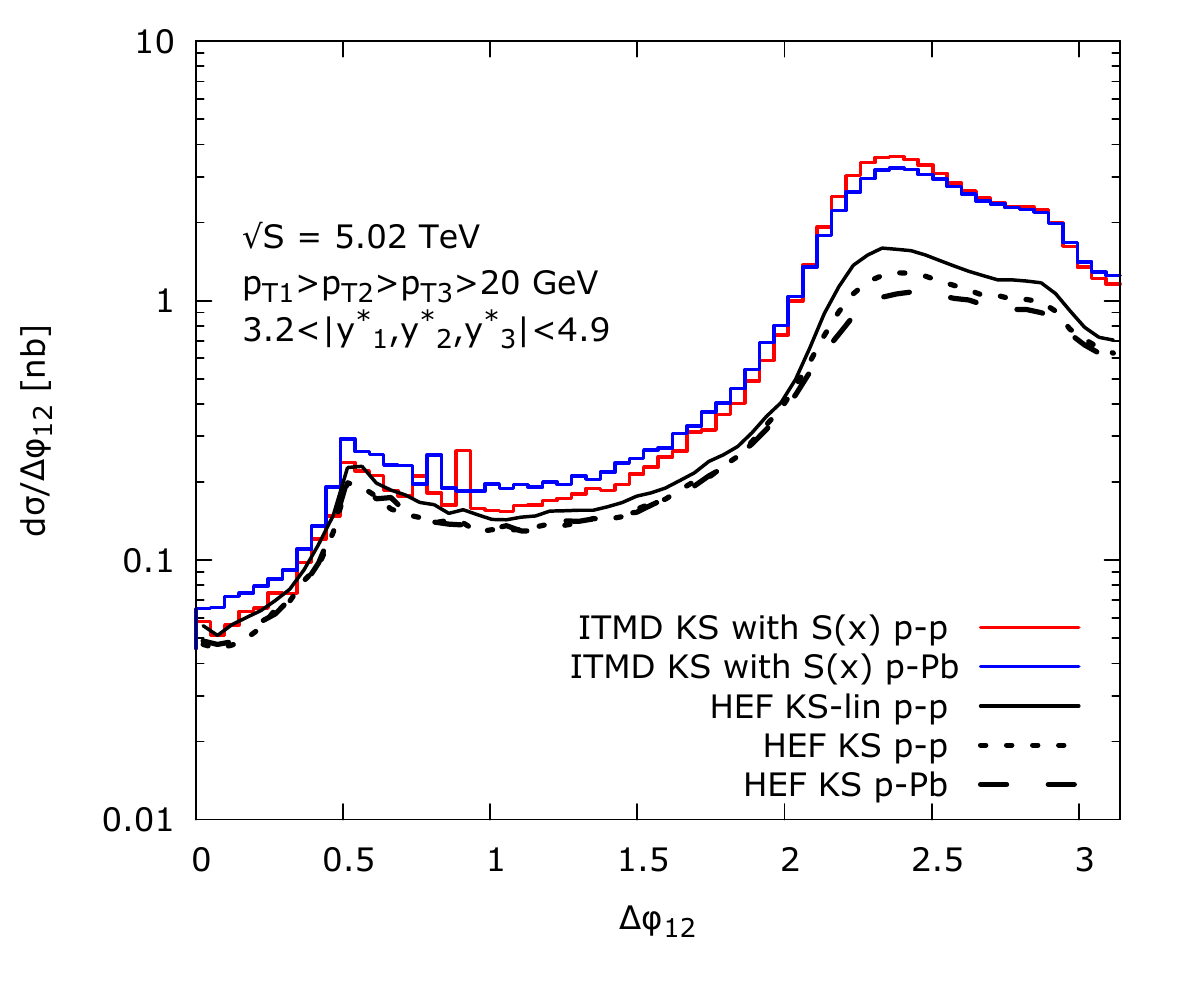}
\end{center}
\end{minipage}
   \caption{\label{fig:phi12}Differential cross sections in the azimuthal angle between the two hardest jets $\Delta\phi_{12}$ for p-p and p-Pb collisions. Left plot represents calculation with uncertainty due to scale variation. Right plot shows the comparison of the ITMD and HEF for central values.}
\end{figure}

\begin{figure}
\begin{minipage}{14.5cm}
\begin{center}
\includegraphics[width=7cm]{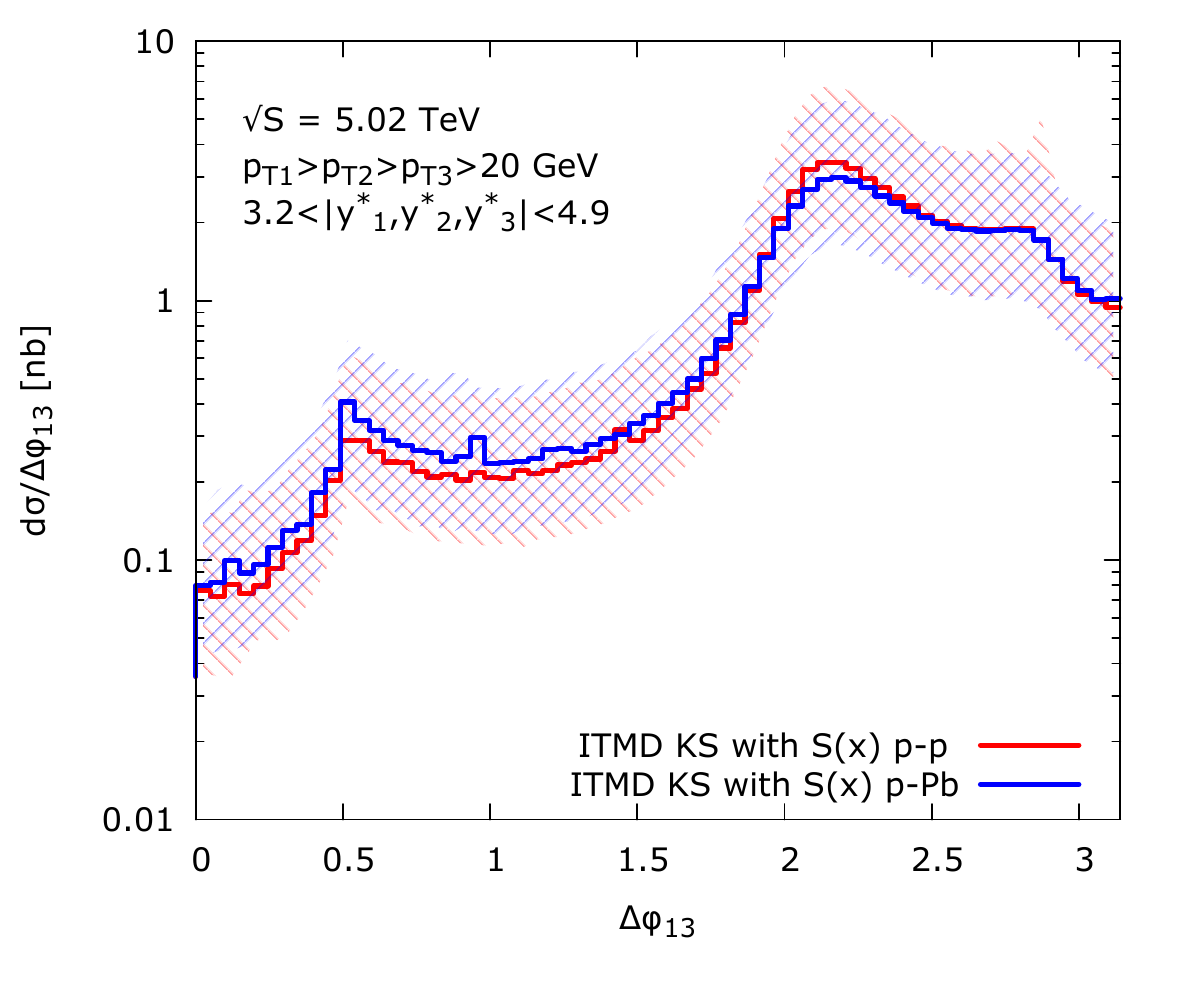}
\includegraphics[width=7cm]{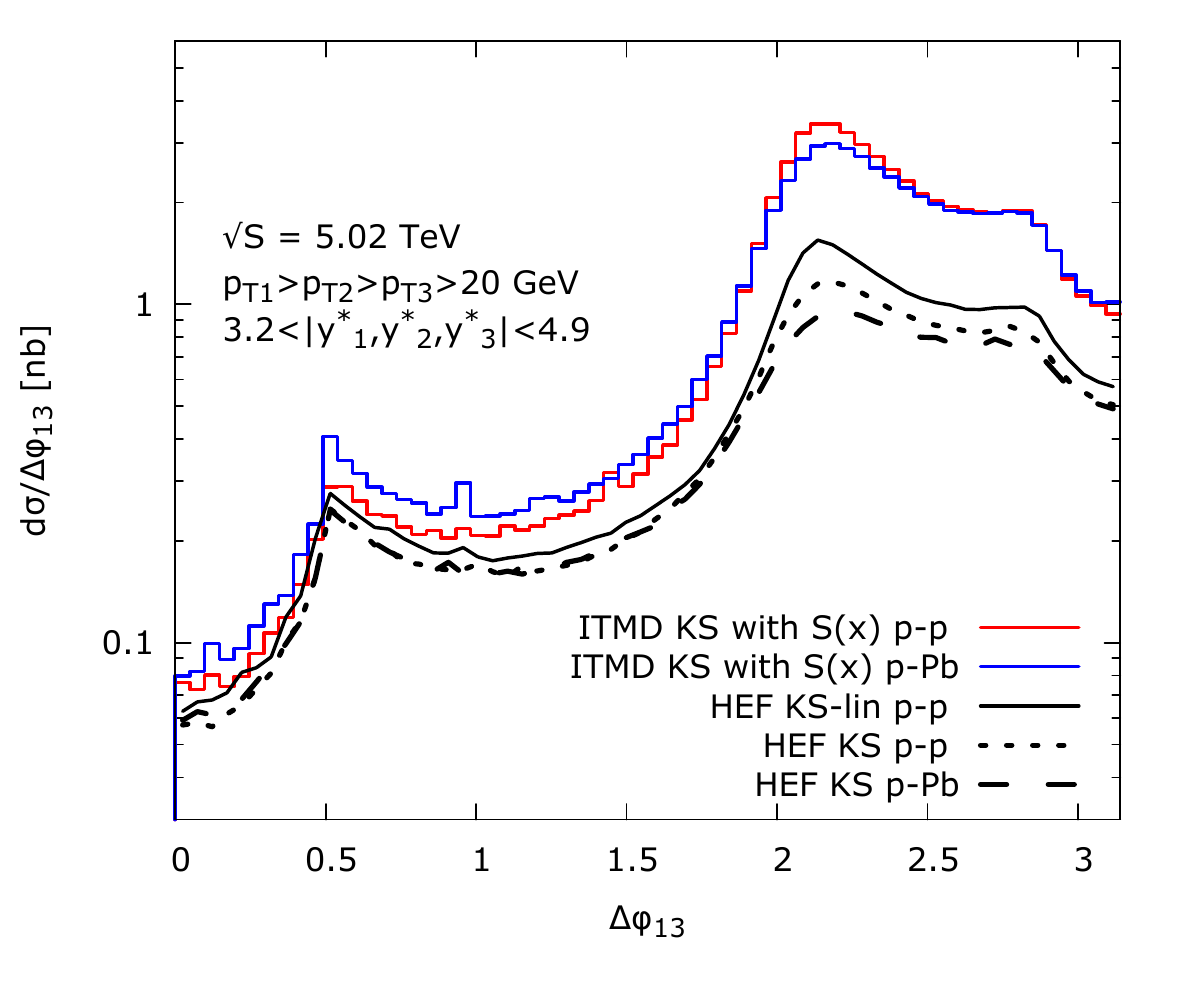}
\end{center}
\end{minipage}
   \caption{\label{fig:phi13}Differential cross sections in the azimuthal angle between the the hardest jet and the softest jet, $\Delta\phi_{13}$, for p-p and p-Pb collisions. Left plot represents the calculation with uncertainty due to scale variation. Right plot shows the comparison of the ITMD and HEF for central values.}
\end{figure}

\begin{figure}
\begin{minipage}{14.5cm}
\begin{center}
\includegraphics[width=7cm]{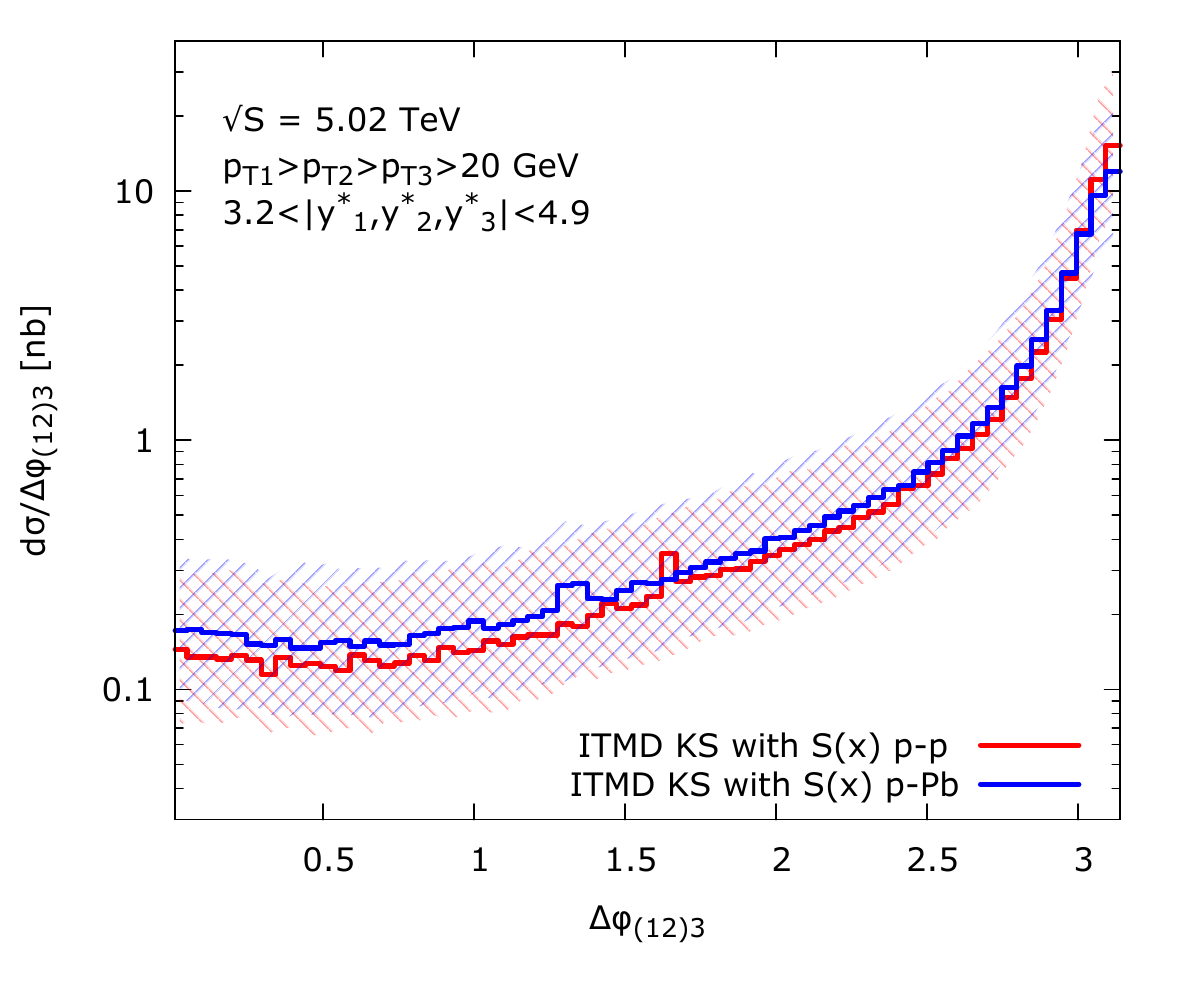}
\includegraphics[width=7cm]{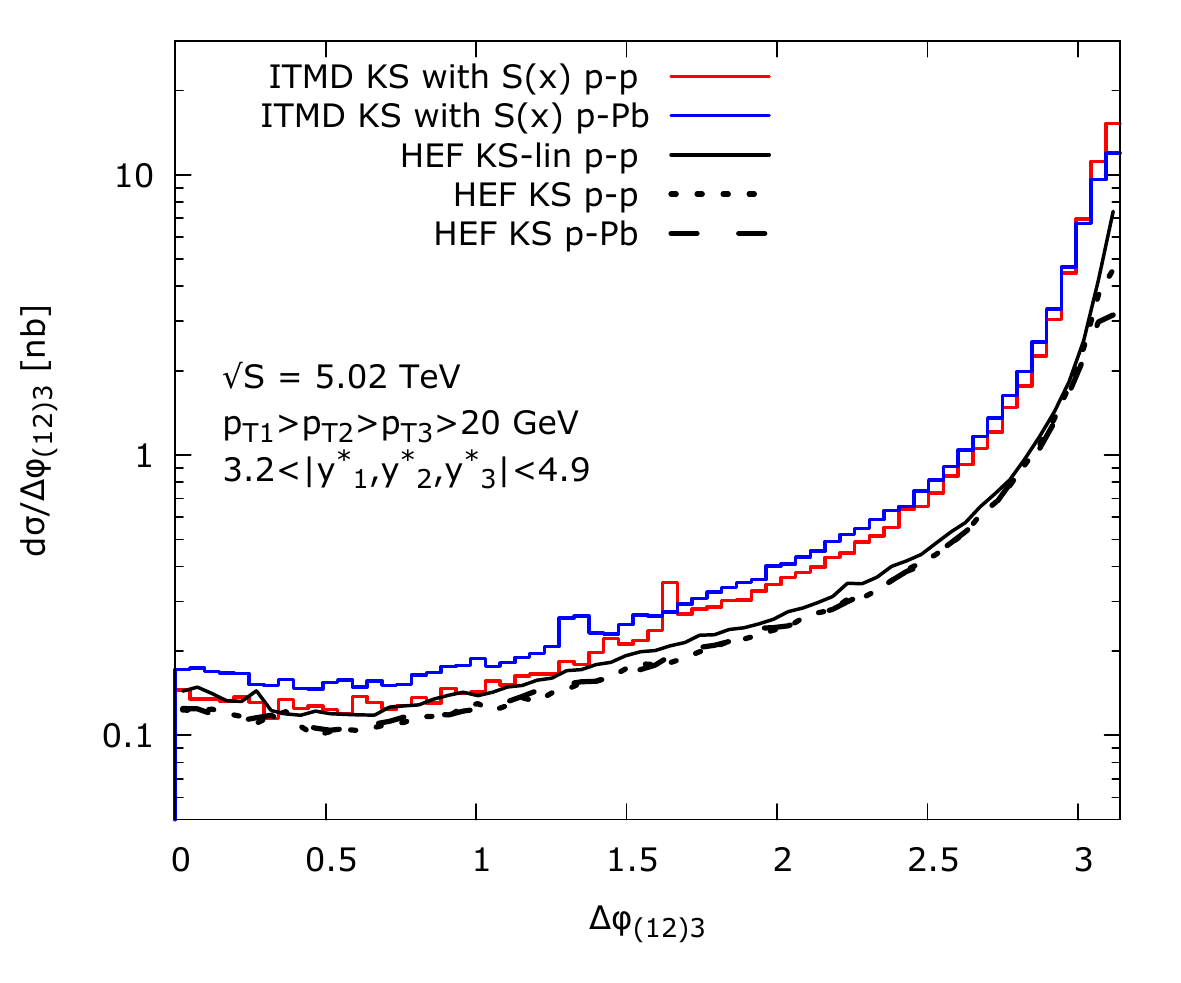}
\end{center}
\end{minipage}
   \caption{\label{fig:phi123}Differential cross sections in the azimuthal angle between the system made of two leading jets $(12)$ and the third jet, $\Delta\phi_{(12)3}$, for p-p and p-Pb collisions. Top plots represent calculation with uncertainty due to scale variation.Left plot represents the calculation with uncertainty due to scale variation. Right plot shows the comparison of the ITMD and HEF for central values.}
\end{figure}

From the panels in Fig.~\ref{fig:RpA} we see  that in all considered scenarios (except the fully dilute limit, for which the ratio would be one) a deviation from unity is clearly visible. 
Especially sensitive is $\Delta\phi_{(12)3}$, which shows a significant suppression in the back-to-back region, indicating strong saturation effects.
Furthermore, we see that for some scenarios $R_{\mathrm{pPb}}$ exceeds unity towards smaller values of azimuthal angles. We link this behavior to already discussed properties of ITMD* gluons where the ratio ${\cal F}_{Pb}/{\cal F}_{p}$ exceeds unity.
In Figs.~\ref{fig:phi12}, \ref{fig:phi13}, \ref{fig:phi123} we plot the absolute cross section, differential in $\Delta\phi_{12}$, $\Delta\phi_{13}$, and $\Delta\phi_{(12)3}$. 
We see that in addition to the already discussed suppression and enhancement of the p-Pb cross section (per nucleon), the normalisation of ITMD* is significantly larger than for HEF in the correlation region.
We attribute this feature to a visibly different shape and larger normalization of the TMD gluons not present in the HEF formalism. Indeed, as seen from Figs.~\ref{fig:KSTMDx}, \ref{fig:Fqg_ratios} they start to dominate over the dipole gluon density as one enters the saturation region.
It is clearly visible in Fig.~\ref{fig:KSTMDx} once we compare for instance $F_{qg}^{(2)}$ and the dipole distribution $F_{qg}^{(1)}$.
Finally, let us note that the sharp peak at $\Delta\phi=0.5$ is a relic of the singularity regularized by the jet algorithm. A simulation with a  proper parton shower and hadronisation would smooth the peak. At present, such modules are not available for the saturation framework, in particular for ITMD (for a recent progress in matching HEF and parton shower see eg. \cite{Bury:2017jxo}).

\section{Conclusions}
In this work, we have presented calculations within an extension of small-$x$ Improved TMD factorization, which was originally proposed in~\cite{Kotko:2015ura} for two final-state partons, to the case of three final-state partons, and designated it ITMD*.
Our extension takes into account gauge invariant hard matrix elements involving off-shell eikonally-coupled initial state gluons, split into several nonequivalent color flows and corresponding TMD gluon distributions.%
At leading order, the formalism can be used to calculate trijet production in forward rapidity region in p-p and p-Pb collisions.
The construction relies  on an application of correlators obtained in \cite{Bury:2018kvg} together with the sets of gauge invariant hard coefficient  that match the color correlators.

Using the ITMD* factorization we calculated various azimuthal-angle-related observables, both for p-p and p-Pb, as well as related nuclear modification ratios $R_{\mathrm{pPb}}$.
We observe significant saturation effects, visible especially in $R_{\mathrm{pPb}}$ as a function of the azimuthal angle difference between the plane spanned by two leading jets and the third jet.
In addition, our results show that there is a significant difference between results obtained using the ITMD* and the standard $k_T$-factorization/high energy factorization (HEF). 
First of all, the ITMD* results give higher cross sections in the correlation region compared to HEF, which is visible in all absolute cross section plots. 
 Since the differences are rather large  we expect them to be a good discriminator of theoretical frameworks.
Secondly, the ITMD* result is not bound to unity, which for the present set of TMD gluon distributions is a consequence of broadening of the $k_T$ distribution for large $k_T$ and moderate $x$. 

In the end, let us stress that ITMD 
factorization is a consequence of saturation effects. If there is no saturation, then the ITMD cross section formula reduces to HEF and all the TMD distributions reduce to a single gluon density being a solution to the linear evolution equation. 
In that sense, the differences between the ITMD* and HEF frameworks we observe reflect the proper account for saturation effects, and thus provide an excellent discrimination tool.

\section{Acknowledgements}
The authors are grateful to C. Marquet for useful comments and H. Van Heavermaet, P. Van Mechelen, M. Pieters, P. Taels for discussions.
 KK and AvH thank for support by the European Union’s Horizon 2020 research and innovation programme under grant agreement No.~824093. PK acknowledges partial support by Narodowe Centrum Nauki grant DEC-2017/27/B/ST2/01985.
\appendix
\newcommand{\Appendix}[1]{Appendix~\ref{#1}}   
\newcommand{\Section}[1]{Section~\ref{#1}}
\newcommand{\Table}[1]{Table~\ref{#1}}
\newcommand{\Figure}[1]{Fig.~\ref{#1}}
\newcommand{\Equation}[1]{Eq.~(\ref{#1})}
\newcommand{\ie}{{\it i.e.}}
\newcommand{\cf}{{\it cf.}}
\newcommand{\eg}{{\it e.g.}}
\newcommand{\Tr}{\mathrm{Tr}}
\newcommand{\imag}{\mathrm{i}}
\newcommand{\PartAmp}{\mathcal{A}}
\newcommand{\FulAmp}{\mathcal{M}}
\newcommand{\Fpdf}{\mathcal{F}}
\newcommand{\kTfactorization}{$k_T$-factorization}
\newcommand{\pdf}{pdf}
\newcommand{\colorMat}{\mathcal{C}}
\newcommand{\Nc}{N_{c}}
\newcommand{\bari}{\bar{\imath}}
\newcommand{\barj}{\bar{\jmath}}
\newcommand{\UU}[2] {\Big(\mathcal{U}^{[{#1}]}\Big)_{#2}}
\newcommand{\UUd}[2]{\Big(\mathcal{U}^{[{#1}]\dagger}\Big)^{#2}}
\newcommand{\UF}[3]{\Big(\hat{F}^{+}(#1)\Big)^{#2}_{#3}}
\newcommand{\matF}{\mathcal{F}}
\newcommand{\VSPACEa}{\vspace{0.5ex}}
\newcommand{\VSPACEb}{\vspace{0.8ex}}
\newcommand{\HSa}{\hspace{-1.5ex}}
\newcommand{\HSb}{\hspace{-1.5ex}}
\section{\label{AppendixA}ITMD factorization in the color connection representation}
The factorization formula for hybrid \kTfactorization\ involves integrals over kinematical variables and an integrand including the factors
%
\begin{equation}
\Fpdf\,|\FulAmp|^2
~,
\label{AVH2}
\end{equation}
%
where $\Fpdf$ is the $k_T$-dependent \pdf, and $|\FulAmp|^2$ is the matrix element of the hard scattering process.
The essential difference in the ITMD factorization formula involves these two factors.
In order to illustrate this, it is useful to represent the matrix element in the color connection representation of~\cite{Kanaki:2000ms,Papadopoulos:2005ky}, because it leads to particularly transparent formulas for ITMD factorization.
This is also the color representation employed in \KaTie.
The matrix element involves a sum over all color degrees of freedom of the external particles in the hard process.
For a hard process with $n_g$ external gluons and $n_q$ external quark-antiquark pairs, the squared amplitude implies
%
\begin{equation}
|\FulAmp|^2 = \sum_{a_1,\ldots,a_{n_g}}\sum_{i_1,\ldots,i_{n_q}}\sum_{j_1,\ldots,j_{n_q}}
\left(
\FulAmp^{a_1\cdots a_{n_g}\;i_1\cdots i_{n_q}}_{\phantom{a_1\cdots a_{n_g}}\;j_1\cdots j_{n_q}}
\right)^*
\left(
\FulAmp^{a_1\cdots a_{n_g}\;i_1\cdots i_{n_q}}_{\phantom{a_1\cdots a_{n_g}}\;j_1\cdots j_{n_q}}
\right)
~,
\end{equation}
%
where $a_1,\ldots,a_{n_g}$ are the adjoint color indices of the gluons, $i_1,\ldots,i_{n_q}$ are the fundamental color indices of the quarks, and $j_1,\ldots,j_{n_q}$ are those of the antiquarks.
The color connection representation is obtained by introducing fundamental color indices for the gluons through contracting every adjoint index $a$ with $(\sqrt{2}\,T^{a})^k_l$, where the $T^a$ are the $SU(\Nc)$ generators, so
%
\begin{equation}
\FulAmp^{a_1\cdots a_{n_g}\;i_1\cdots i_{n_q}}_{\phantom{a_1\cdots a_{n_g}}\;j_1\cdots j_{n_q}}
\;\;\to\;\;
\tilde{\FulAmp}^{k_1\cdots k_{n_g}\;i_1\cdots i_{n_q}}_{l_1\cdots l_{n_g}\;j_1\cdots j_{n_q}}
\;=
\sum_{a_1,\ldots,a_{n_g}}
\FulAmp^{a_1\cdots a_{n_g}\;i_1\cdots i_{n_q}}_{\phantom{a_1\cdots a_{n_g}}\;j_1\cdots j_{n_q}}\,\big(\sqrt{2}\,T^{a_1}\big)^{k_1}_{l_1}\cdots\big(\sqrt{2}\,T^{a_{n_g}}\big)^{k_{n_g}}_{l_{n_g}}
~.
\end{equation}
%
Because of the identity
%
\begin{equation}
\delta^{ab} =
2\mathrm{Tr}\{T^aT^b\} =
\sum_{i,j}\big(\big(\sqrt{2}\,T^b\big)^i_j)^*\big(\sqrt{2}\,T^a\big)^i_j
~,
\end{equation}
%
the matrix element can now be written as
%
\begin{equation}
|\FulAmp|^2 = 
\sum_{k_1,\ldots,k_{n_g}}\sum_{l_1,\ldots,l_{n_g}}\sum_{i_1,\ldots,i_{n_q}}\sum_{j_1,\ldots,j_{n_q}}
\left(
\tilde{\FulAmp}^{k_1\cdots k_{n_g}\;i_1\cdots i_{n_q}}_{l_1\cdots l_{n_g}\;j_1\cdots j_{n_q}}
\right)^*
\left(
\tilde{\FulAmp}^{k_1\cdots k_{n_g}\;i_1\cdots i_{n_q}}_{l_1\cdots l_{n_g}\;j_1\cdots j_{n_q}}
\right)
~.
\label{AVH1}
\end{equation}
%
In~\cite{Papadopoulos:2005ky} it is explained how $\tilde{\FulAmp}$ can be calculated directly with adjusted Feynman rules regarding color.
Notice that the unavoidable ``$1/\Nc$-correction'' is in the quark-gluon vertex rather than in the gluon propagator like in~\cite{Maltoni:2002mq}, avoiding the need for projectors in~\Equation{AVH1}.

From now on we will use the same symbol $i$ for color-indices of gluons and quarks, and $j$ for anti-color indices for gluons and anti-quarks, and write $n=n_g+n_q$.
The scattering amplitude $\tilde{\FulAmp}$ can be decomposed into color factors and partial amplitudes following
%
\begin{equation}
\tilde{\FulAmp}^{i_1 i_2\cdots i_{n}}_{j_1 j_2\cdots j_{n}}
=
\sum_{\sigma\in S_{n}}
\delta^{i_{1}}_{j_{\sigma(1)}}
\delta^{i_{2}}_{j_{\sigma(2)}}
\cdots
\delta^{i_{n}}_{j_{\sigma(n)}}
\,
\PartAmp_\sigma
\,
\label{AVH3}
\end{equation}
%
were $S_n$ is the group of all permutations of $(1,2,\ldots,n)$, and where the partial amplitudes $\PartAmp_\sigma$ do not depend on color, but may include factors of $1/\Nc$.
If the scattering process does not involve quark-antiquark pairs, then $\PartAmp_\sigma$ actually vanishes for many permutations, which can be expressed explicitly with formula (3) in~\cite{Maltoni:2002mq}.
The formula above, however, holds for any process.

Inserting \Equation{AVH3} into \Equation{AVH1}, the matrix element can be expressed in terms of partial amplitudes via a color matrix
%
\begin{equation}
|\FulAmp|^2 = \sum_{\sigma\in S_n}\sum_{\tau\in S_n}
\PartAmp_\sigma^*\colorMat_{\sigma\tau}\PartAmp_\tau
~,
\end{equation}
with
\begin{equation}
\colorMat_{\sigma\tau}
=
\sum_{i_1,\ldots,i_{n}}\sum_{j_1,\ldots,j_{n}}
\delta^{i_{1}}_{j_{\sigma(1)}}
\delta^{i_{2}}_{j_{\sigma(2)}}
\cdots
\delta^{i_{n}}_{j_{\sigma(n)}}
\times
\delta^{i_{1}}_{j_{\tau(1)}}
\delta^{i_{2}}_{j_{\tau(2)}}
\cdots
\delta^{i_{n}}_{j_{\tau(n)}}
~.
\end{equation}
%
It is not difficult to see that each entry of the matrix $\colorMat_{\sigma\tau}$ consists of a single power of $\Nc$.

Despite the fact that one can simply ignore vanishing partial amplitudes, the formula is still not optimal from the point of view of computational efficiency, since the partial amplitudes are linearly dependent.
This is exploited in the color representation of~\cite{DelDuca:1999rs}, leading to the smallest possible color matrices, but with more complicated entries consisting of polynomials in $\Nc$.
Below we will stick to the color representation with the big matrices with simple entries.

Let us assume that the off-shell gluon is the one carrying label number $1$.
In ITMD factorization, \Equation{AVH2} is replaced with
%
\begin{multline}
\matF\,|\FulAmp|^2 \;\;\to\;\; (\Nc^2-1)
\sum_{i_1,\ldots,i_{n}}\sum_{j_1,\ldots,j_{n}}\sum_{\bari_1,\ldots,\bari_{n}}\sum_{\barj_1,\ldots,\barj_{n}}
\left(
\tilde{\FulAmp}^{i_1i_2\cdots i_{n}}_{j_1j_2\cdots j_{n}}
\right)^*
\left(
\tilde{\FulAmp}^{\bari_1\bari_2\cdots\bari_{n}}_{\barj_1\barj_2\cdots\barj_{n}}
\right)
\\\times
\bigg\langle\!\!\bigg\langle
2\UF{\xi}{j_1}{i_1}\UF{0}{\barj_1}{\bari_1}\;
\UU{\lambda_2}{i_2\bari_2}\UUd{\lambda_2}{j_2\barj_2}
\cdots
\UU{\lambda_n}{i_n\bari_n}\UUd{\lambda_n}{j_n\barj_n}
\bigg\rangle\!\!\bigg\rangle
~,
\label{AVH4}
\end{multline}
%
where the symbols ~$\mathcal{U}^{[\lambda]}$ denote the Wilson lines of \Equation{eq:staple}, with $\lambda=\pm$ depending on whether the parton whose color it connects is incoming or outgoing, and $\hat{F}^+$ is the field strength.
The double brackets represent both the hadronic matrix element and the Fourier transform.
Notice that the formula does not distinguish between gluons and quark-antiquark pairs, and that compared to the formulas in Table~2 of~\cite{Bury:2018kvg}, the ``$1/\Nc$-terms'' are absent.
Now, we can insert \Equation{AVH3} again, and find that the right-hand side of \Equation{AVH4} can be written as
%
\begin{equation}
(\Nc^2-1)
\sum_{\sigma\in S_n}\sum_{\tau\in S_n}
\PartAmp_\sigma^*\hat{\colorMat}_{\sigma\tau}\PartAmp_\tau
~,
\end{equation}
%
where the entries of the ``TMD-valued'' color matrix $\hat{\colorMat}_{\sigma\tau}$ consist exactly of a single power of $\Nc$ times one of the $10$ $\mathcal{F}$-functions from \Equation{eq:Fqg1} to \Equation{eq:Fgg7}.
The relevant matrices for $3$-jet production are given in \Table{TMDtable1}, \Table{TMDtable2}, \Table{TMDtable3}, and \Table{TMDtable4}.
%

\begin{table}
\begin{center}
$$
\left(
\begin{array}{cccccc}
     0 &                   0 &                   0 &                   0 &                   0 & 0 \VSPACEa\\
     0 & \Nc\matF_{qg}^{(1)} &    \matF_{qg}^{(1)} &                   0 &    \matF_{qg}^{(1)} & 0 \VSPACEa\\
     0 &    \matF_{qg}^{(1)} & \Nc\matF_{qg}^{(2)} &    \matF_{qg}^{(3)} &                   0 & 0 \VSPACEa\\
     0 &                   0 &    \matF_{qg}^{(3)} & \Nc\matF_{qg}^{(2)} &    \matF_{qg}^{(1)} & 0 \VSPACEa\\
     0 &    \matF_{qg}^{(1)} &                   0 &    \matF_{qg}^{(1)} & \Nc\matF_{qg}^{(1)} & 0 \VSPACEa\\
     0 &                   0 &                   0 &                   0 &                   0 & 0 
\end{array}
\right)
\left(
\begin{array}{c}
     \PartAmp_{12345} \VSPACEb\\
     \PartAmp_{21345} \VSPACEb\\
     \PartAmp_{23145} \VSPACEb\\
     \PartAmp_{32145} \VSPACEb\\
     \PartAmp_{31245} \VSPACEb\\
     \PartAmp_{13245} 
\end{array}
\right)
$$
\caption{\label{TMDtable1}%
TMD-valued color matrix and vector of partial amplitudes for the processes $g^*_1\,q_2 \to q_4\,\bar{q}'_3\,q'_5$ and $g^*_1\,q_2 \to q_4\,\bar{q}_3\,q_5$.
The $6$ partial amplitudes are explicitly labeled with their associated permutation. The logic in the enumation of the partons is that gluons come first, and then anti-quarks, where initial-state quarks count as negative-energy antiquarks. For clarity, all 6 partial amplitudes are included, also the non-contributing ones.}
\end{center}
\end{table}

\begin{table}
\begin{center}
\scriptsize
\begin{tabular}{cccccccccccccccccr}
$22$&$\HSa12$&$\HSa 0$&$\HSa12$&$\HSa01$&$\HSa 0$&$\HSa01$&$\HSa 0$&$\HSa11$&$\HSa 0$&$\HSa03$&$\HSa13$&$\HSa03$&$\HSa03$&$\HSa12$&$\HSa01$&$\HSa 0$&$21345$\\
$12$&$\HSa22$&$\HSa12$&$\HSa 0$&$\HSa11$&$\HSa01$&$\HSa 0$&$\HSa 0$&$\HSa01$&$\HSa03$&$\HSa13$&$\HSa03$&$\HSa 0$&$\HSa 0$&$\HSa01$&$\HSa12$&$\HSa03$&$23145$\\
$ 0$&$\HSa12$&$\HSa22$&$\HSa12$&$\HSa01$&$\HSa11$&$\HSa01$&$\HSa 0$&$\HSa 0$&$\HSa12$&$\HSa03$&$\HSa 0$&$\HSa01$&$\HSa03$&$\HSa 0$&$\HSa03$&$\HSa13$&$32145$\\
$12$&$\HSa 0$&$\HSa12$&$\HSa22$&$\HSa 0$&$\HSa01$&$\HSa11$&$\HSa 0$&$\HSa01$&$\HSa01$&$\HSa 0$&$\HSa03$&$\HSa12$&$\HSa13$&$\HSa03$&$\HSa 0$&$\HSa03$&$31245$\\
$01$&$\HSa11$&$\HSa01$&$\HSa 0$&$\HSa21$&$\HSa11$&$\HSa01$&$\HSa01$&$\HSa11$&$\HSa 0$&$\HSa 0$&$\HSa 0$&$\HSa 0$&$\HSa01$&$\HSa11$&$\HSa01$&$\HSa 0$&$23415$\\
$ 0$&$\HSa01$&$\HSa11$&$\HSa01$&$\HSa11$&$\HSa21$&$\HSa11$&$\HSa11$&$\HSa01$&$\HSa01$&$\HSa 0$&$\HSa01$&$\HSa11$&$\HSa 0$&$\HSa 0$&$\HSa 0$&$\HSa 0$&$32415$\\
$01$&$\HSa 0$&$\HSa01$&$\HSa11$&$\HSa01$&$\HSa11$&$\HSa21$&$\HSa01$&$\HSa11$&$\HSa11$&$\HSa01$&$\HSa 0$&$\HSa01$&$\HSa 0$&$\HSa 0$&$\HSa 0$&$\HSa 0$&$34215$\\
$ 0$&$\HSa 0$&$\HSa 0$&$\HSa 0$&$\HSa01$&$\HSa11$&$\HSa01$&$\HSa21$&$\HSa11$&$\HSa 0$&$\HSa01$&$\HSa11$&$\HSa01$&$\HSa01$&$\HSa 0$&$\HSa01$&$\HSa11$&$42315$\\
$11$&$\HSa01$&$\HSa 0$&$\HSa01$&$\HSa11$&$\HSa01$&$\HSa11$&$\HSa11$&$\HSa21$&$\HSa 0$&$\HSa 0$&$\HSa 0$&$\HSa 0$&$\HSa 0$&$\HSa01$&$\HSa11$&$\HSa01$&$24315$\\
$ 0$&$\HSa03$&$\HSa12$&$\HSa01$&$\HSa 0$&$\HSa01$&$\HSa11$&$\HSa 0$&$\HSa 0$&$\HSa22$&$\HSa12$&$\HSa 0$&$\HSa12$&$\HSa 0$&$\HSa03$&$\HSa13$&$\HSa03$&$34125$\\
$03$&$\HSa13$&$\HSa03$&$\HSa 0$&$\HSa 0$&$\HSa 0$&$\HSa01$&$\HSa01$&$\HSa 0$&$\HSa12$&$\HSa22$&$\HSa12$&$\HSa 0$&$\HSa01$&$\HSa 0$&$\HSa03$&$\HSa12$&$43125$\\
$13$&$\HSa03$&$\HSa 0$&$\HSa03$&$\HSa 0$&$\HSa01$&$\HSa 0$&$\HSa11$&$\HSa 0$&$\HSa 0$&$\HSa12$&$\HSa22$&$\HSa12$&$\HSa12$&$\HSa03$&$\HSa 0$&$\HSa01$&$41325$\\
$03$&$\HSa 0$&$\HSa01$&$\HSa12$&$\HSa 0$&$\HSa11$&$\HSa01$&$\HSa01$&$\HSa 0$&$\HSa12$&$\HSa 0$&$\HSa12$&$\HSa22$&$\HSa03$&$\HSa13$&$\HSa03$&$\HSa 0$&$31425$\\
$03$&$\HSa 0$&$\HSa03$&$\HSa13$&$\HSa01$&$\HSa 0$&$\HSa 0$&$\HSa01$&$\HSa 0$&$\HSa 0$&$\HSa01$&$\HSa12$&$\HSa03$&$\HSa22$&$\HSa12$&$\HSa 0$&$\HSa12$&$41235$\\
$12$&$\HSa01$&$\HSa 0$&$\HSa03$&$\HSa11$&$\HSa 0$&$\HSa 0$&$\HSa 0$&$\HSa01$&$\HSa03$&$\HSa 0$&$\HSa03$&$\HSa13$&$\HSa12$&$\HSa22$&$\HSa12$&$\HSa 0$&$21435$\\
$01$&$\HSa12$&$\HSa03$&$\HSa 0$&$\HSa01$&$\HSa 0$&$\HSa 0$&$\HSa01$&$\HSa11$&$\HSa13$&$\HSa03$&$\HSa 0$&$\HSa03$&$\HSa 0$&$\HSa12$&$\HSa22$&$\HSa12$&$24135$\\
$ 0$&$\HSa03$&$\HSa13$&$\HSa03$&$\HSa 0$&$\HSa 0$&$\HSa 0$&$\HSa11$&$\HSa01$&$\HSa03$&$\HSa12$&$\HSa01$&$\HSa 0$&$\HSa12$&$\HSa 0$&$\HSa12$&$\HSa22$&$42135$
\end{tabular}
\caption{\label{TMDtable2}%
Representation of the TMD-valued color matrix and vector of partial amplitudes for the processes $g^*_1\,q_4 \to g_2\,g_3\,q_5$.
Each pair $ij$ of intergers represents $\Nc^i\,\matF_{qg}^{(j)}$, and a single $0$ means that the entry vanishes.
The last column gives the permutation associated with the partial amplitudes.
}
\end{center}
\end{table}
\begin{table}
\begin{center}
\scriptsize
\begin{tabular}{cccccccccccccccccr}
%
%
$24$&$\HSa14$&$\HSa 0$&$\HSa14$&$\HSa04$&$\HSa 0$&$\HSa04$&$\HSa 0$&$\HSa14$&$\HSa 0$&$\HSa04$&$\HSa14$&$\HSa04$&$\HSa04$&$\HSa14$&$\HSa04$&$\HSa 0$&$21345$\\
$14$&$\HSa21$&$\HSa13$&$\HSa22$&$\HSa11$&$\HSa03$&$\HSa12$&$\HSa 0$&$\HSa04$&$\HSa03$&$\HSa11$&$\HSa04$&$\HSa12$&$\HSa12$&$\HSa04$&$\HSa11$&$\HSa03$&$23145$\\
$ 0$&$\HSa13$&$\HSa23$&$\HSa13$&$\HSa03$&$\HSa13$&$\HSa03$&$\HSa 0$&$\HSa 0$&$\HSa13$&$\HSa03$&$\HSa 0$&$\HSa03$&$\HSa03$&$\HSa 0$&$\HSa03$&$\HSa13$&$32145$\\
$14$&$\HSa22$&$\HSa13$&$\HSa21$&$\HSa12$&$\HSa03$&$\HSa11$&$\HSa 0$&$\HSa04$&$\HSa03$&$\HSa12$&$\HSa04$&$\HSa11$&$\HSa11$&$\HSa04$&$\HSa12$&$\HSa03$&$31245$\\
$04$&$\HSa11$&$\HSa03$&$\HSa12$&$\HSa21$&$\HSa13$&$\HSa03$&$\HSa03$&$\HSa11$&$\HSa12$&$\HSa 0$&$\HSa12$&$\HSa22$&$\HSa04$&$\HSa14$&$\HSa04$&$\HSa 0$&$23415$\\
$ 0$&$\HSa03$&$\HSa13$&$\HSa03$&$\HSa13$&$\HSa23$&$\HSa13$&$\HSa13$&$\HSa03$&$\HSa03$&$\HSa 0$&$\HSa03$&$\HSa13$&$\HSa 0$&$\HSa 0$&$\HSa 0$&$\HSa 0$&$32415$\\
$04$&$\HSa12$&$\HSa03$&$\HSa11$&$\HSa03$&$\HSa13$&$\HSa26$&$\HSa03$&$\HSa11$&$\HSa17$&$\HSa05$&$\HSa12$&$\HSa03$&$\HSa 0$&$\HSa 0$&$\HSa 0$&$\HSa 0$&$34215$\\
$ 0$&$\HSa 0$&$\HSa 0$&$\HSa 0$&$\HSa03$&$\HSa13$&$\HSa03$&$\HSa23$&$\HSa13$&$\HSa 0$&$\HSa03$&$\HSa13$&$\HSa03$&$\HSa03$&$\HSa 0$&$\HSa03$&$\HSa13$&$42315$\\
$14$&$\HSa04$&$\HSa 0$&$\HSa04$&$\HSa11$&$\HSa03$&$\HSa11$&$\HSa13$&$\HSa21$&$\HSa 0$&$\HSa12$&$\HSa22$&$\HSa12$&$\HSa12$&$\HSa04$&$\HSa11$&$\HSa03$&$24315$\\
$ 0$&$\HSa03$&$\HSa13$&$\HSa03$&$\HSa12$&$\HSa03$&$\HSa17$&$\HSa 0$&$\HSa 0$&$\HSa26$&$\HSa17$&$\HSa 0$&$\HSa11$&$\HSa12$&$\HSa04$&$\HSa11$&$\HSa03$&$34125$\\
$04$&$\HSa11$&$\HSa03$&$\HSa12$&$\HSa 0$&$\HSa 0$&$\HSa05$&$\HSa03$&$\HSa12$&$\HSa17$&$\HSa26$&$\HSa11$&$\HSa 0$&$\HSa03$&$\HSa 0$&$\HSa03$&$\HSa13$&$43125$\\
$14$&$\HSa04$&$\HSa 0$&$\HSa04$&$\HSa12$&$\HSa03$&$\HSa12$&$\HSa13$&$\HSa22$&$\HSa 0$&$\HSa11$&$\HSa21$&$\HSa11$&$\HSa11$&$\HSa04$&$\HSa12$&$\HSa03$&$41325$\\
$04$&$\HSa12$&$\HSa03$&$\HSa11$&$\HSa22$&$\HSa13$&$\HSa03$&$\HSa03$&$\HSa12$&$\HSa11$&$\HSa 0$&$\HSa11$&$\HSa21$&$\HSa04$&$\HSa14$&$\HSa04$&$\HSa 0$&$31425$\\
$04$&$\HSa12$&$\HSa03$&$\HSa11$&$\HSa04$&$\HSa 0$&$\HSa 0$&$\HSa03$&$\HSa12$&$\HSa12$&$\HSa03$&$\HSa11$&$\HSa04$&$\HSa21$&$\HSa14$&$\HSa22$&$\HSa13$&$41235$\\
$14$&$\HSa04$&$\HSa 0$&$\HSa04$&$\HSa14$&$\HSa 0$&$\HSa 0$&$\HSa 0$&$\HSa04$&$\HSa04$&$\HSa 0$&$\HSa04$&$\HSa14$&$\HSa14$&$\HSa24$&$\HSa14$&$\HSa 0$&$21435$\\
$04$&$\HSa11$&$\HSa03$&$\HSa12$&$\HSa04$&$\HSa 0$&$\HSa 0$&$\HSa03$&$\HSa11$&$\HSa11$&$\HSa03$&$\HSa12$&$\HSa04$&$\HSa22$&$\HSa14$&$\HSa21$&$\HSa13$&$24135$\\
$ 0$&$\HSa03$&$\HSa13$&$\HSa03$&$\HSa 0$&$\HSa 0$&$\HSa 0$&$\HSa13$&$\HSa03$&$\HSa03$&$\HSa13$&$\HSa03$&$\HSa 0$&$\HSa13$&$\HSa 0$&$\HSa13$&$\HSa23$&$42135$
\end{tabular}
\caption{\label{TMDtable3}%
Representation of the TMD-valued color matrix and vector of partial amplitudes for the processes $g^*_1\,g_2 \to g_3\,\bar{q}_4\,q_5$.
Each pair $ij$ of intergers represents $\Nc^i\,\matF_{gg}^{(j)}$, and a single $0$ means that the entry vanishes.
The last column gives the permutation associated with the partial amplitudes.
}
\end{center}
\end{table}

\begin{table}
\begin{center}
\scriptsize
\begin{tabular}{ccccccccccccccccccccccccc}
$\HSb31$&$\HSa13$&$\HSa11$&$\HSa13$&$\HSa11$&$\HSa11$&$\HSa12$&$\HSa 0$&$\HSa12$&$\HSa12$&$\HSa32$&$\HSa 0$&$\HSa 0$&$\HSa14$&$\HSa14$&$\HSa14$&$\HSa 0$&$\HSa14$&$\HSa12$&$\HSa11$&$\HSa11$&$\HSa12$&$\HSa11$&$\HSa12$&$23451$\\
$\HSb13$&$\HSa36$&$\HSa17$&$\HSa13$&$\HSa17$&$\HSa11$&$\HSa17$&$\HSa 0$&$\HSa17$&$\HSa12$&$\HSa13$&$\HSa15$&$\HSa 0$&$\HSa 0$&$\HSa 0$&$\HSa 0$&$\HSa 0$&$\HSa 0$&$\HSa11$&$\HSa12$&$\HSa12$&$\HSa11$&$\HSa12$&$\HSa11$&$34251$\\
$\HSb11$&$\HSa17$&$\HSa36$&$\HSa17$&$\HSa13$&$\HSa13$&$\HSa15$&$\HSa17$&$\HSa 0$&$\HSa13$&$\HSa12$&$\HSa17$&$\HSa11$&$\HSa11$&$\HSa12$&$\HSa11$&$\HSa12$&$\HSa12$&$\HSa 0$&$\HSa 0$&$\HSa 0$&$\HSa 0$&$\HSa 0$&$\HSa 0$&$43521$\\
$\HSb13$&$\HSa13$&$\HSa17$&$\HSa36$&$\HSa11$&$\HSa11$&$\HSa17$&$\HSa15$&$\HSa12$&$\HSa12$&$\HSa13$&$\HSa 0$&$\HSa 0$&$\HSa 0$&$\HSa 0$&$\HSa 0$&$\HSa 0$&$\HSa 0$&$\HSa12$&$\HSa11$&$\HSa12$&$\HSa17$&$\HSa17$&$\HSa11$&$35421$\\
$\HSb11$&$\HSa17$&$\HSa13$&$\HSa11$&$\HSa36$&$\HSa13$&$\HSa 0$&$\HSa12$&$\HSa15$&$\HSa13$&$\HSa12$&$\HSa17$&$\HSa17$&$\HSa12$&$\HSa12$&$\HSa11$&$\HSa17$&$\HSa11$&$\HSa 0$&$\HSa 0$&$\HSa 0$&$\HSa 0$&$\HSa 0$&$\HSa 0$&$45231$\\
$\HSb11$&$\HSa11$&$\HSa13$&$\HSa11$&$\HSa13$&$\HSa31$&$\HSa 0$&$\HSa12$&$\HSa 0$&$\HSa32$&$\HSa12$&$\HSa12$&$\HSa12$&$\HSa12$&$\HSa11$&$\HSa12$&$\HSa11$&$\HSa11$&$\HSa14$&$\HSa14$&$\HSa14$&$\HSa 0$&$\HSa 0$&$\HSa14$&$24531$\\
$\HSb12$&$\HSa17$&$\HSa15$&$\HSa17$&$\HSa 0$&$\HSa 0$&$\HSa36$&$\HSa17$&$\HSa 0$&$\HSa 0$&$\HSa11$&$\HSa17$&$\HSa12$&$\HSa12$&$\HSa11$&$\HSa12$&$\HSa11$&$\HSa11$&$\HSa 0$&$\HSa 0$&$\HSa13$&$\HSa13$&$\HSa 0$&$\HSa13$&$34512$\\
$\HSb 0$&$\HSa 0$&$\HSa17$&$\HSa15$&$\HSa12$&$\HSa12$&$\HSa17$&$\HSa36$&$\HSa11$&$\HSa11$&$\HSa 0$&$\HSa 0$&$\HSa 0$&$\HSa13$&$\HSa 0$&$\HSa 0$&$\HSa13$&$\HSa13$&$\HSa11$&$\HSa12$&$\HSa11$&$\HSa17$&$\HSa17$&$\HSa12$&$53412$\\
$\HSb12$&$\HSa17$&$\HSa 0$&$\HSa12$&$\HSa15$&$\HSa 0$&$\HSa 0$&$\HSa11$&$\HSa36$&$\HSa 0$&$\HSa11$&$\HSa17$&$\HSa17$&$\HSa11$&$\HSa11$&$\HSa12$&$\HSa17$&$\HSa12$&$\HSa13$&$\HSa13$&$\HSa 0$&$\HSa 0$&$\HSa13$&$\HSa 0$&$54132$\\
$\HSb12$&$\HSa12$&$\HSa13$&$\HSa12$&$\HSa13$&$\HSa32$&$\HSa 0$&$\HSa11$&$\HSa 0$&$\HSa31$&$\HSa11$&$\HSa11$&$\HSa11$&$\HSa11$&$\HSa12$&$\HSa11$&$\HSa12$&$\HSa12$&$\HSa14$&$\HSa14$&$\HSa14$&$\HSa 0$&$\HSa 0$&$\HSa14$&$41532$\\
$\HSb32$&$\HSa13$&$\HSa12$&$\HSa13$&$\HSa12$&$\HSa12$&$\HSa11$&$\HSa 0$&$\HSa11$&$\HSa11$&$\HSa31$&$\HSa 0$&$\HSa 0$&$\HSa14$&$\HSa14$&$\HSa14$&$\HSa 0$&$\HSa14$&$\HSa11$&$\HSa12$&$\HSa12$&$\HSa11$&$\HSa12$&$\HSa11$&$31452$\\
$\HSb 0$&$\HSa15$&$\HSa17$&$\HSa 0$&$\HSa17$&$\HSa12$&$\HSa17$&$\HSa 0$&$\HSa17$&$\HSa11$&$\HSa 0$&$\HSa36$&$\HSa13$&$\HSa 0$&$\HSa13$&$\HSa13$&$\HSa 0$&$\HSa 0$&$\HSa12$&$\HSa11$&$\HSa11$&$\HSa12$&$\HSa11$&$\HSa12$&$43152$\\
$\HSb 0$&$\HSa 0$&$\HSa11$&$\HSa 0$&$\HSa17$&$\HSa12$&$\HSa12$&$\HSa 0$&$\HSa17$&$\HSa11$&$\HSa 0$&$\HSa13$&$\HSa36$&$\HSa 0$&$\HSa13$&$\HSa13$&$\HSa15$&$\HSa 0$&$\HSa12$&$\HSa11$&$\HSa12$&$\HSa17$&$\HSa17$&$\HSa11$&$45123$\\
$\HSb14$&$\HSa 0$&$\HSa11$&$\HSa 0$&$\HSa12$&$\HSa12$&$\HSa12$&$\HSa13$&$\HSa11$&$\HSa11$&$\HSa14$&$\HSa 0$&$\HSa 0$&$\HSa31$&$\HSa14$&$\HSa14$&$\HSa13$&$\HSa32$&$\HSa11$&$\HSa12$&$\HSa12$&$\HSa12$&$\HSa11$&$\HSa11$&$51423$\\
$\HSb14$&$\HSa 0$&$\HSa12$&$\HSa 0$&$\HSa12$&$\HSa11$&$\HSa11$&$\HSa 0$&$\HSa11$&$\HSa12$&$\HSa14$&$\HSa13$&$\HSa13$&$\HSa14$&$\HSa31$&$\HSa32$&$\HSa 0$&$\HSa14$&$\HSa12$&$\HSa11$&$\HSa11$&$\HSa12$&$\HSa11$&$\HSa12$&$24153$\\
$\HSb14$&$\HSa 0$&$\HSa11$&$\HSa 0$&$\HSa11$&$\HSa12$&$\HSa12$&$\HSa 0$&$\HSa12$&$\HSa11$&$\HSa14$&$\HSa13$&$\HSa13$&$\HSa14$&$\HSa32$&$\HSa31$&$\HSa 0$&$\HSa14$&$\HSa11$&$\HSa12$&$\HSa12$&$\HSa11$&$\HSa12$&$\HSa11$&$41253$\\
$\HSb 0$&$\HSa 0$&$\HSa12$&$\HSa 0$&$\HSa17$&$\HSa11$&$\HSa11$&$\HSa13$&$\HSa17$&$\HSa12$&$\HSa 0$&$\HSa 0$&$\HSa15$&$\HSa13$&$\HSa 0$&$\HSa 0$&$\HSa36$&$\HSa13$&$\HSa11$&$\HSa12$&$\HSa11$&$\HSa17$&$\HSa17$&$\HSa12$&$54213$\\
$\HSb14$&$\HSa 0$&$\HSa12$&$\HSa 0$&$\HSa11$&$\HSa11$&$\HSa11$&$\HSa13$&$\HSa12$&$\HSa12$&$\HSa14$&$\HSa 0$&$\HSa 0$&$\HSa32$&$\HSa14$&$\HSa14$&$\HSa13$&$\HSa31$&$\HSa12$&$\HSa11$&$\HSa11$&$\HSa11$&$\HSa12$&$\HSa12$&$25413$\\
$\HSb12$&$\HSa11$&$\HSa 0$&$\HSa12$&$\HSa 0$&$\HSa14$&$\HSa 0$&$\HSa11$&$\HSa13$&$\HSa14$&$\HSa11$&$\HSa12$&$\HSa12$&$\HSa11$&$\HSa12$&$\HSa11$&$\HSa11$&$\HSa12$&$\HSa31$&$\HSa32$&$\HSa14$&$\HSa 0$&$\HSa13$&$\HSa14$&$51234$\\
$\HSb11$&$\HSa12$&$\HSa 0$&$\HSa11$&$\HSa 0$&$\HSa14$&$\HSa 0$&$\HSa12$&$\HSa13$&$\HSa14$&$\HSa12$&$\HSa11$&$\HSa11$&$\HSa12$&$\HSa11$&$\HSa12$&$\HSa12$&$\HSa11$&$\HSa32$&$\HSa31$&$\HSa14$&$\HSa 0$&$\HSa13$&$\HSa14$&$25134$\\
$\HSb11$&$\HSa12$&$\HSa 0$&$\HSa12$&$\HSa 0$&$\HSa14$&$\HSa13$&$\HSa11$&$\HSa 0$&$\HSa14$&$\HSa12$&$\HSa11$&$\HSa12$&$\HSa12$&$\HSa11$&$\HSa12$&$\HSa11$&$\HSa11$&$\HSa14$&$\HSa14$&$\HSa31$&$\HSa13$&$\HSa 0$&$\HSa32$&$23514$\\
$\HSb12$&$\HSa11$&$\HSa 0$&$\HSa17$&$\HSa 0$&$\HSa 0$&$\HSa13$&$\HSa17$&$\HSa 0$&$\HSa 0$&$\HSa11$&$\HSa12$&$\HSa17$&$\HSa12$&$\HSa12$&$\HSa11$&$\HSa17$&$\HSa11$&$\HSa 0$&$\HSa 0$&$\HSa13$&$\HSa36$&$\HSa15$&$\HSa13$&$35214$\\
$\HSb11$&$\HSa12$&$\HSa 0$&$\HSa17$&$\HSa 0$&$\HSa 0$&$\HSa 0$&$\HSa17$&$\HSa13$&$\HSa 0$&$\HSa12$&$\HSa11$&$\HSa17$&$\HSa11$&$\HSa11$&$\HSa12$&$\HSa17$&$\HSa12$&$\HSa13$&$\HSa13$&$\HSa 0$&$\HSa15$&$\HSa36$&$\HSa 0$&$53124$\\
$\HSb12$&$\HSa11$&$\HSa 0$&$\HSa11$&$\HSa 0$&$\HSa14$&$\HSa13$&$\HSa12$&$\HSa 0$&$\HSa14$&$\HSa11$&$\HSa12$&$\HSa11$&$\HSa11$&$\HSa12$&$\HSa11$&$\HSa12$&$\HSa12$&$\HSa14$&$\HSa14$&$\HSa32$&$\HSa13$&$\HSa 0$&$\HSa31$&$31524$
\end{tabular}
\caption{\label{TMDtable4}%
Representation of the TMD-valued color matrix and vector of partial amplitudes for the processes $g^*_1\,g_2 \to g_3\,g_4\,g_5$.
Each pair $ij$ of intergers represents $\Nc^i\,\matF_{gg}^{(j)}$, and a single $0$ means that the entry vanishes.
The last column gives the permutation associated with the partial amplitudes.
}
\end{center}
\end{table}

\bibliographystyle{JHEP}
\bibliography{references}

\end{document}